\documentclass[twocolumn,preprintnumbers,amsmath,amssymb,aps,pre,reprint,longbibliography]{revtex4-1}
\usepackage{graphicx}
\usepackage{xcolor}
\begin{document}

\title{Peak Effect and Dynamics of Stripe and Pattern Forming Systems on a Periodic One Dimensional Substrate
}
\author{
C. Reichhardt and C. J. O. Reichhardt 
} 
\affiliation{
Theoretical Division and Center for Nonlinear Studies,
Los Alamos National Laboratory, Los Alamos, New Mexico 87545, USA
}

\date{\today}

\begin{abstract}
We examine the ordering, pinning, and dynamics of two-dimensional pattern forming systems interacting with a periodic one-dimensional substrate. In the absence of the substrate, particles with competing long-range repulsion and short-range attraction form anisotropic crystal, stripe, and bubble states. When the system is tuned across the stripe transition in the presence of a substrate, we find that there is a peak effect in the critical depinning force when the stripes align and become commensurate with the substrate. Under an applied drive, the anisotropic crystal and stripe states can exhibit soliton depinning and plastic flow. When the stripes depin plastically, they dynamically reorder into a moving stripe state that is perpendicular to the substrate trough direction. We also find that when the substrate spacing is smaller than the widths of the bubbles or stripes, the system forms pinned stripe states that are perpendicular to the substrate trough direction. The system exhibits multiple reentrant pinning effects as a function of increasing attraction, with the anisotropic crystal and large bubble states experiencing weak pinning but the stripe and smaller bubble states showing stronger pinning. We map out the different dynamic phases as a function of filling, the strength of the attractive interaction term, the substrate strength, and the drive, and demonstrate that the different phases produce identifiable features in the transport curves and particle orderings.
\end{abstract}

\maketitle

\section{Introduction}

Particle systems with competing long-range repulsive and short-range attractive
interactions form a variety of patterned states, including
crystals, stripes, bubbles, and void lattices
\cite{Seul95,Stoycheva00,Reichhardt03,Reichhardt04,Mossa04,Sciortino04,Nelissen05,Liu08,Reichhardt10,McDermott16,Liu19,Xu21}.
For a fixed repulsion strength but increasing attraction
strength, these systems
form a crystal, then an anisotropic crystal,
and finally stripe and bubble phases \cite{Reichhardt10}.
For fixed interaction strength but increasing particle densities,
first bubbles, then stripes, then void lattices, and then a uniform crystal
appear \cite{Reichhardt10}.
Similar patterns can arise even for systems with purely
repulsive interactions if the interaction potential involves multiple
length scales
\cite{Jagla98,Malescio03,Glaser07}.
Pattern formation can occur in soft matter systems
such as colloidal assemblies,
emulsions, and binary fluids \cite{Malescio03,Glaser07,CostaCampos13},
and in hard
condensed matter systems that include electron liquid crystals
\cite{Fogler96,Moessner96,Cooper99,Pan99,Fradkin99,Gores07,Zhu09,Friess18},
composite fermion states \cite{Shingla23},
multiple component superconducting vortex systems
\cite{Xu11,Komendova13,Varney13,Sellin13,Brems22},
skyrmion systems \cite{Reichhardt22a},
and various types of charge ordered states
\cite{Tranquada95,Reichhardt04a,Mertelj05,Baity18,Mahmoudian15}.
When pattern forming systems couple to 
quenched disorder,
they can exhibit pinned phases as well as depinning
transitions and sliding phases
under an applied drive
\cite{Cooper99,Reichhardt03,Reichhardt03a,Gores07,Zhu09,Zhao13,Reichhardt17,Bennaceur18,Friess18,Brems22,Sun22}.
If the quenched
disorder is strong, a glassy or structurally disordered state
forms that depins plastically, and for high drives
the system can dynamically reorder into patterned states
such as moving stripes or moving bubbles
\cite{Reichhardt03,Reichhardt03a,Gores07,Sun22}.
The different dynamic states and transitions between them
are associated with multiple steps in the
transport curves
\cite{Cooper99,Reichhardt03,Reichhardt03a,Gores07,Reichhardt17},
changes in the noise fluctuations
\cite{Reichhardt03a,Qian17,Bennaceur18,Sun22,Madathil23},
and modifications of the structure factors \cite{Reichhardt03a,Brems22}.

There have been extensive studies of systems of purely repulsive particles
that form crystalline lattices under coupling to
one- or two-dimensional periodic substrates
\cite{Chowdhury85,Chakrabarti95,Harada96,Frey99,Wei98,Bechinger01,Reichhardt02a,Brunner02,Reichhardt17}.
Far less is known about how a pattern-forming system with competing
interactions would behave when coupled to a periodic substrate.
For particle systems with purely repulsive interactions, such as certain types
of colloidal particles \cite{Chowdhury85,Bechinger01,Brunner02}
and superconducting vortices
\cite{Martinoli78,Harada96,Reichhardt97,LeThien16,Berdiyorov06},
the relevant length scales are the average spacing between the particles and
the periodicity of the substrate.
In contrast, for
stripe or bubble forming systems, 
additional length scales arise including 
the spacing between adjacent stripes or bubbles as well as
the average spacing between the particles that compose each stripe or
bubble,
so a richer variety
of commensuration effects are possible.
Additionally, the mesoscale morphology in pattern-forming
systems permits the appearance
of matching or pinning effects that are not possible for repulsive
point particles.
For example, a stripe might show strong commensuration effects when interacting
with a periodic one-dimensional (1D) substrate
since the stripe can easily match the substrate shape.
In general, if the attraction or repulsion strength or the filling fraction
of the system is varied,
morphologies can emerge that are more strongly pinned due to better matching
with the substrate length scales or shape,
while for other morphologies, the patterns do not match,
leading to changes in the pinning configurations, sliding, and transport.

Previous work on the static configurations of pattern forming systems
on a periodic 1D substrate identified
several new types of patterns,
such as modulated stripes and anisotropic bubbles \cite{McDermott14}.
Recently, we studied the depinning of bubbles on periodic
1D substrates under a dc drive applied
parallel to the substrate periodicity direction,
and found that the bubbles can depin either elastically or
plastically depending on the
substrate strength \cite{Reichhardt24}.
In addition, as the strength of the attractive interaction term
increased, the bubbles became smaller 
and better pinned since they could
fit within the substrate minima better than larger bubbles.
When the bubbles depin
elastically, there is a single peak in the differential velocity-force curves,
while for plastic depinning,
multiple peaks appear when the bubbles break up and move in various modes,
such as via individual particles hopping from bubble to bubble or
via a moving bubble shedding individual particles.
At higher drives,
the system can dynamically reorder
back into a moving bubble lattice through a transition
similar to the dynamic ordering found
for superconducting vortices, colloidal particles,
Wigner crystals, and skyrmions moving over random substrates
\cite{Bhattacharya93,Moon96,Balents98,Olson98a,delaCruz98,Reichhardt15,Reichhardt17}.

In this work, we consider the pinning and dynamics of a
pattern forming system with competing long-range repulsion
and short-range attraction
interacting with a 1D periodic substrate as we sweep through
parameters where
crystal, stripe, and bubble states appear in a clean system.
We find that the depinning threshold shows a peak or
maximum in the stripe regime
when the stripes form a commensurate state that aligns with the substrate
troughs.
When the substrate is strong, the stripes first depin
plastically via the formation of
running kinks or solitons,  followed by the emergence of a
disordered flowing state,
while at high drives, there
is a dynamical transition into a moving stripe phase
where the stripes are wider and rotate with respect to
the pinned configuration so that they are aligned with the
driving direction.
The anisotropic
crystal state can also exhibit soliton depinning, disordered motion,
and dynamical reordering into an
anisotropic crystal at high drives.
The drive needed to induce the reordering transition
diverges near the boundary between the stripe and anisotropic
crystal states.
When the substrate lattice constant is decreased,
the stripe and anisotropic crystal states
remain strongly pinned, but the bubble phases show a pronounced
depinning threshold decrease.
For constant drives that are well above the depinning threshold,
the average velocity passes through
a dip in the stripe phase.
When the substrate lattice constant is considerably smaller than the width
of the stripes or bubbles,
the system forms a pinned stripe or modulated stripe aligned perpendicular
to the substrate trough direction that can
depin into moving stripes or moving bubbles.
When the particle-particle interaction strengths are held fixed while
the filling fraction of the system is changed,
we find that all three phases show step-like features in the
depinning threshold that correlate with
commensuration effects in which an integer number of rows of particles
can fit inside an individual substrate trough.
At high filling fractions, the stripes depin
into a modulated solid that remains aligned with the substrate trough
direction even at high drives.
We also find that as the substrate strength increases,
there is a sharp increase in the depinning threshold at the
crossover from elastic to plastic depinning.
For strong pinning at a fixed drive,
the velocity is a non-monotonic function of the
magnitude of the attractive interaction term.
The velocities are highest at zero attraction, pass through
a minimum or reentrant pinning region
in the stripe state, increase again for large bubbles,
and then decrease until a second reentrant pinning regime for
small bubbles emerges.

\section{Simulation}

We consider a two-dimensional (2D) system
with periodic boundary conditions in the $x-$ and $y$-directions.
The sample
contains $N$ particles that have pairwise interactions composed
of a long-range repulsive term, which favors
formation of a uniform triangular lattice,
and a competing short-range attractive term,
which favors clump or bubble formation.
In our system the repulsion dominates at very short distances, which
prevents complete collapse of the particles to a point
even for strong attractive interactions.
The system is of size $L \times L$ with $L=36$ and the
particle density is $\rho = N/L^2$.
The dynamics of particle $i$ obey the
following overdamped equation of motion:
\begin{equation}
\eta \frac{d {\bf R}_{i}}{dt} =
-\sum^{N}_{j \neq i} \nabla V(R_{ij}) + {\bf F}^{s}_{i} +
        {\bf F}_{D} ,
\end{equation}
where the damping term is set to
$\eta=1.0$.
The first term on the right hand side describes
the particle-particle interactions,
where
\begin{equation}
V(R_{ij}) = \frac{1}{R_{ij}} - B\exp(-\kappa R_{ij}) \ .
\end{equation}
Here
$R_{ij}=|{\bf R}_i-{\bf R}_j|$,
and the location of particle $i (j)$ is
${\bf R}_{i (j)}$.
For computational efficiency, we treat the
long range
repulsive Coulomb interaction using a
real space Lekner summation technique
as in previous work \cite{Reichhardt03,McDermott14}.
The short range attraction term falls off exponentially
with distance. The interaction potential of Eq.~(2)
produces crystal, stripe, bubble, or void lattice
states depending on the values of $\rho$, the attractive
force strength $B$, and the inverse screening length $\kappa$
\cite{Reichhardt03,Reichhardt03a,Reichhardt04,Reichhardt10,McDermott14}.
In this work we fix $\kappa = 1.0$. We
focus on a
particle density of $\rho = 0.44$, but 
also consider a range of densities from
$\rho=0.01$ to $\rho=1.2$.
For $\rho = 0.44$ in the absence of substrate,
the system forms a crystal for $B < 2.0$, a stripe state
for $2.0 \leq B < 2.25$, and bubbles for $B \geq 2.25$.

The second term on the right hand side of Eq.~(2) represents
the interaction with
a 1D substrate that is sinusoidal in form
with $N_p$ minima and a lattice constant of $a_{p} = L/N_{p}$.
Here
\begin{equation}
{\bf F}_s^i = F_p \cos(2\pi x_i/a_p){\bf \hat{x}}
\end{equation}
where $x_i$ is the $x$ position of particle $i$.
We focus on substrates with $N_{p} = 8$,
but also consider $N_p=4$, 17, and $35$.
The particles are subjected to a uniform driving
force ${\bf F}_D=F_D {\bf {\hat{x}}}$.

The initial particle configuration is obtained through simulated
annealing
by placing
the particles in a lattice configuration,
subjecting them to a high temperature, and then slowly cooling the
system. The thermal
forces are represented by Langevin kicks, and after the simulated annealing
is complete, the temperature is set to zero.
After initialization, we apply
a driving force
to all of the particles. We typically wait
$10^4$ or more simulation time steps
after changing the driving force to avoid any transient effects,
and then we measure the time-averaged particle velocity in the
driving direction,
$\langle V\rangle = \sum^{N}_i{\bf v}_i\cdot {\hat {\bf x}}$.
From this measure we can construct a velocity-force curve.

\begin{figure}
\includegraphics[width=\columnwidth]{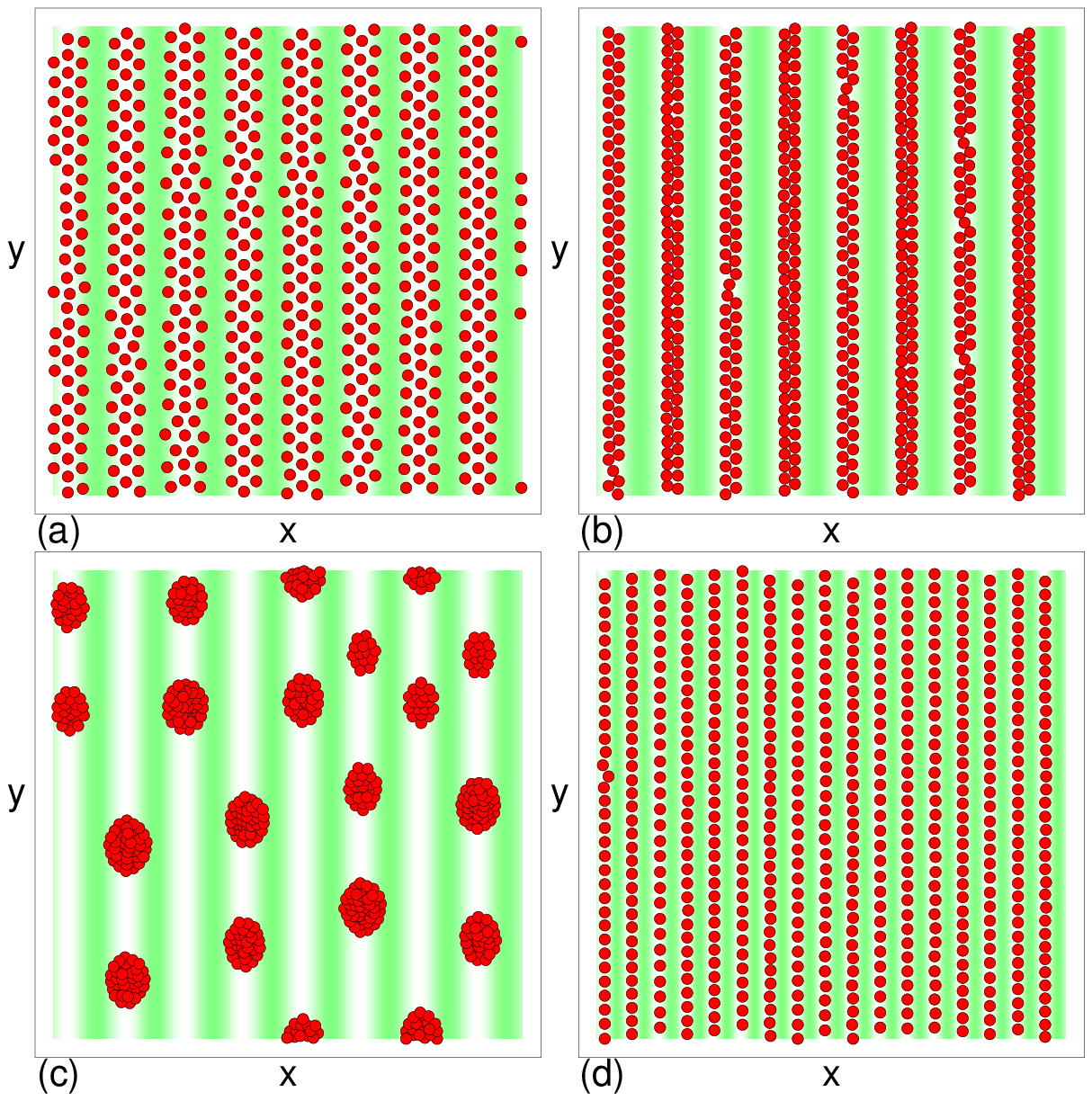}
\caption{(a-c) Particle positions (red circles) and substrate potential
(green shading) for a sample with particle density  
$\rho = 0.44$, pinning strength $F_{p} = 1.0$,
and number of substrate minima $N_{p} = 8$
in the absence of driving, $F_D=0$, for different attractive
interaction strengths $B$.
Note that here and throughout this work, the size of the circles
representing the particles is chosen purely for visual convenience; the
particles do not have a defined outer edge.
(a) An anisotropic crystal at $B = 0.2$.
(b) A stripe state at $B = 2.15$.
(c) A bubble phase at $B = 2.75$.
(d) The $B=2.15$ system from panel (b) with
a higher number of substrate minima, $N_{p} = 17$, where
each stripe is composed of a single row of particles.
} 
\label{fig:1}
\end{figure}

\section{Results}

In Fig.~\ref{fig:1}, we show the pinned particle configurations
at $F_D=0$ for a
system with $\rho = 0.44$,  $N_p = 8$, and $F_{p} = 1.0$. 
At $B=0.2$, in the absence of a substrate a uniform crystal would form,
but as illustrated in Fig.~\ref{fig:1}(a), the presence of a substrate
produces an anisotropic crystal that has small density modulations
induced by the substrate potential.
Figure~\ref{fig:1}(b) shows a stripe phase
at $B = 2.15$, where the stripes are aligned with the substrate troughs and
each trough
contains two rows of particles.
At $B=2.75$, the bubble phase shown in Fig.~\ref{fig:1}(c) forms. 
In Fig.~\ref{fig:1}(d), the stripe state with $B=2.15$ from Fig.~\ref{fig:1}(b)
is placed on a substrate with $N_p=17$, which reduces the substrate lattice
constant. Here, each substrate trough captures a single row of particles.

\begin{figure}
\includegraphics[width=\columnwidth]{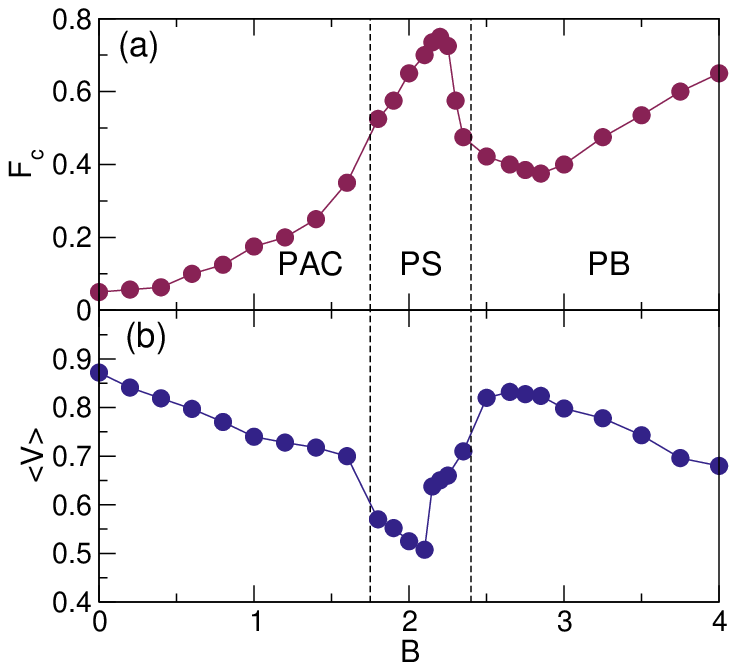}
\caption{(a) The critical depinning force $F_c$ vs $B$
for the system from Fig.~\ref{fig:1}(a,b,c) with
$\rho = 0.44$, $F_{p}= 1.0$, and $N_p=8$.
(b) The average velocity per particle $\langle V\rangle$ vs $B$ at fixed
$F_{D} = 1.0$. The dashed lines indicate that the
system forms a pinned anisotropic crystal (PAC)
for $B < 1.9$, a pinned stripe (PS) state for $1.9 \leq B \leq 2.25$,
and a pinned bubble (PB) state for $B > 2.25$.
In the PS state, there is
a peak in the depinning force and a dip in the velocity.
} 
\label{fig:2} 
\end{figure}

We next examine the driving force $F_c$ at which the system depins
as a function of $B$
for the system in Fig.~\ref{fig:1}(a-c) with $\rho=0.44$, $F_p=1.0$, and
$N_p=8$ by performing a series of simulations and constructing velocity-force
curves.
In Fig.~\ref{fig:2}(a), we plot $F_{c}$ for a range of $B$ values that span
a pinned isotropic crystal (PAC) state,
a pinned stripe (PS) state, and a pinned bubble (PB) state.
For small $B$, the depinning threshold has a low value of
$F_{c} = 0.05$, indicating that the anisotropic crystal phases are weakly pinned.
As $B$ increases, $F_{c}$ increases and reaches a peak value of
$F_c=0.75$ in the stripe phase, showing that the stripes are
strongly pinned.
For $B$ values above the peak in $F_c$, the depinning threshold
decreases with increasing $B$ and
the system enters the bubble phase, where a local minimum of
$F_{c}= 0.39$ appears at $B = 2.9$. As $B$ increases further,
the bubbles shrink in size and
the depinning threshold increases again since the smaller bubbles fit
better into the pinning troughs,
as shown in a previous study \cite{Reichhardt24}.
In Fig.~\ref{fig:2}(b) we plot
the average velocity $\langle V\rangle$ versus $B$ at a constant drive of
$F_{D} = 1.0$.
In the absence of a substrate, $\langle V\rangle = F_D = 1.0$
for this value of $F_{D}$. In the presence of the substrate,
$\langle V\rangle$ decreases with increasing $B$ until it reaches
a local minimum in the stripe phase.
The velocity then increases with increasing $B$ up to a local
maximum value that appears early in the bubble phase, and finally
decreases again with increasing $B$ for large $B$.
If the same measurement is performed at larger values of $F_D$,
we find that the dip in $\langle V\rangle$ in the stripe phase
persists but
becomes flatter with increasing
$F_{D}$.
This dip is a signature of the peak in the
the critical depinning force
that appears when the system
has formed an anisotropic stripe state.

\begin{figure}[h]
\includegraphics[width=\columnwidth]{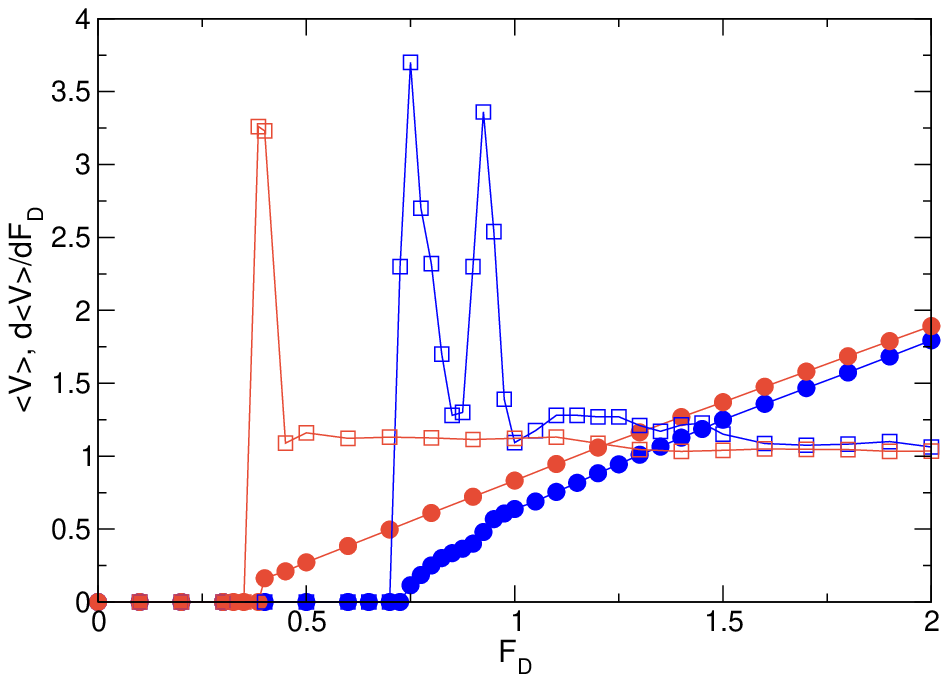}
\caption{
The velocity-force curves $\langle V\rangle$ vs $F_D$ (solid circles)
and the corresponding
differential velocity $d\langle V\rangle/dF_{D}$ vs $F_{D}$ curves
(open squares)
for the system from Fig.~\ref{fig:2} with $\rho = 0.44$, $F_{p} = 1.0$,
and $N_p=8$.
At $B = 2.15$ (blue curves), the system is in the stripe state and there is
a double peak in the differential velocity,
while at $B = 2.75$ (red curves),
the system is in the bubble state and
there is a single peak in $d\langle V\rangle/dF_{D}$.
}
\label{fig:3}
\end{figure}

\begin{figure}
\includegraphics[width=\columnwidth]{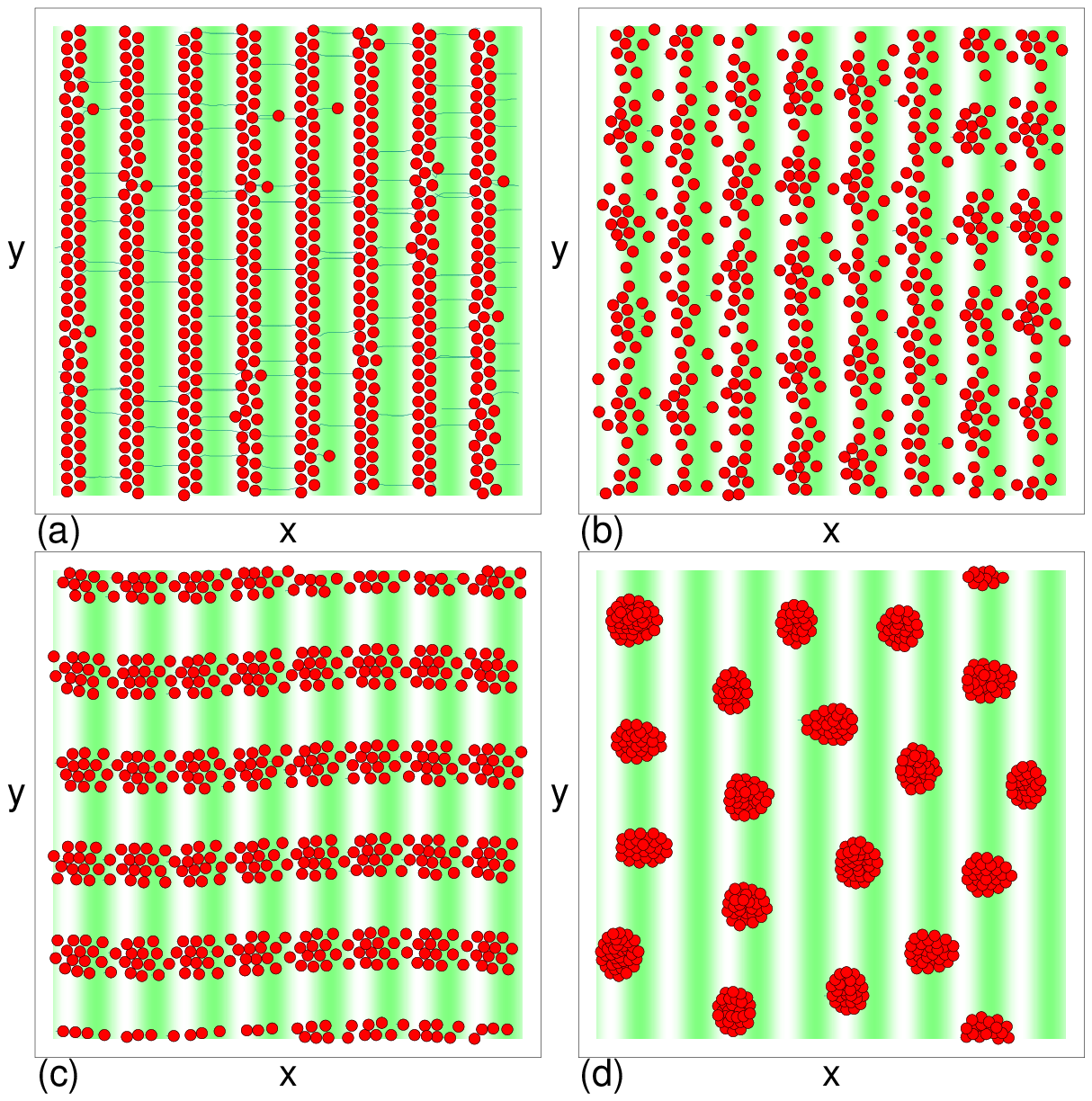}
\caption{Particle positions (red circles) and substrate potential (green
shading) for the system from Fig.~\ref{fig:3} with $\rho=0.44$,
$F_p=1.0$, and $N_p=8.$
(a) The soliton flow phase at $B = 2.15$ and $F_{D} = 0.825$. Lines
indicate the trajectories of individual particles.
(b) A disordered moving phase at $B = 2.15$ and $F_{D} = 1.0$.
(c) A moving stripe phase at $B = 2.15$ and $F_{D} = 1.5$.
(d) A moving bubble phase at $B = 2.75$ and $F_{D}  = 1.5$.
}
\label{fig:5}
\end{figure}

In Fig.~\ref{fig:3} we plot the velocity-force
curves $\langle V\rangle$ along with
the corresponding differential velocity curves
$d\langle V\rangle/dF_{D}$ versus $F_D$
for the system in Fig.~\ref{fig:2} with $\rho=0.44$, $F_p=1.0$, and
$N_p=8$ at $B = 2.75$ in the bubble phase and $B=2.15$ in the stripe phase.
For $B = 2.75$, the depinning threshold is low,
there is a single peak in $d\langle V\rangle/dF_{D}$,
and the differential velocity approaches
$d\langle V\rangle/dF_D=1.0$ just above this peak.
Here, the bubbles depin elastically and pass
directly from a pinned bubble state to a moving bubble phase.
At $B = 2.15$, the stripes depin plastically, and there is
a double peak in the differential conductivity,
with the initial depinning producing a peak near $F_{D}= 0.74$
followed by a second peak in $d\langle V\rangle/dF_D$
near $F_{D}= 0.925$.
The differential velocity does not
approach $d\langle V\rangle/dF_D=1$ until $F_{D} > 1.3$.
The initial depinning of the stripe state
occurs via the sliding of kinks or solitons,
where individual particles hop out of one well and displace a particle
in the adjacent well. This is
illustrated in Fig.~\ref{fig:5}(a) where we
highlight the particle trajectories from the stripe state in
Fig.~\ref{fig:3} at $F_{D}= 0.825$, just above $F_c$.
For drives of $0.95 < F_{D} < 1.2$,
the stripe system forms the moving disordered structure 
shown in Fig.~\ref{fig:5}(b) at $F_{D} = 1.1$,
while at $F_{D} = 1.5$, a moving stripe structure that is aligned with the
driving direction and not the substrate trough direction appears, as
illustrated in Fig.~\ref{fig:5}(c).
Since
the number of stripes in the moving stripe structure is smaller than $N_p$,
each stripe has a width of four particles. 
There is also a periodic density modulation along the length of the stripe.
The moving bubble state that appears above depinning for the
$B=2.75$ system in
Fig.~\ref{fig:3} is illustrated in
Fig.~\ref{fig:5}(d) 
at $F_{D} = 1.5$,
where the bubbles have developed a slight anisotropy favoring
the driving direction.

\begin{figure}
\includegraphics[width=\columnwidth]{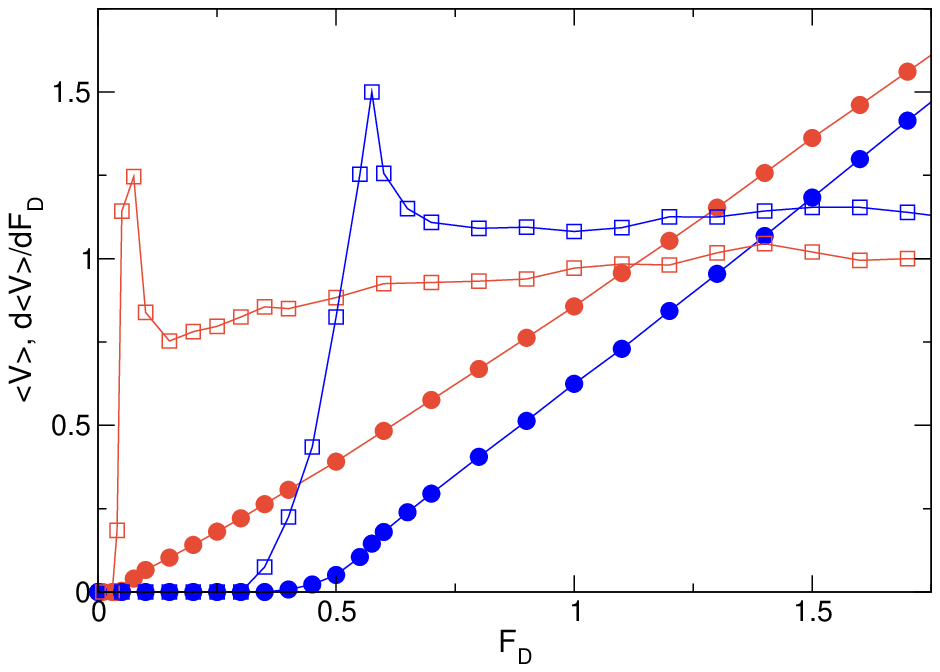}
\caption{$\langle V\rangle$ vs $F_D$ (solid circles) and the corresponding
$d\langle V\rangle/dF_{D}$ vs $F_{D}$ (open squares)
for the system from Fig.~\ref{fig:2} with $\rho = 0.44$, $F_{p} = 1.0$,
and $N_p=8$ in the anisotropic crystal state at $B = 1.6$ (blue curves)
and $B=0.2$ (red curves).
}
\label{fig:4}
\end{figure}

\begin{figure}
\includegraphics[width=\columnwidth]{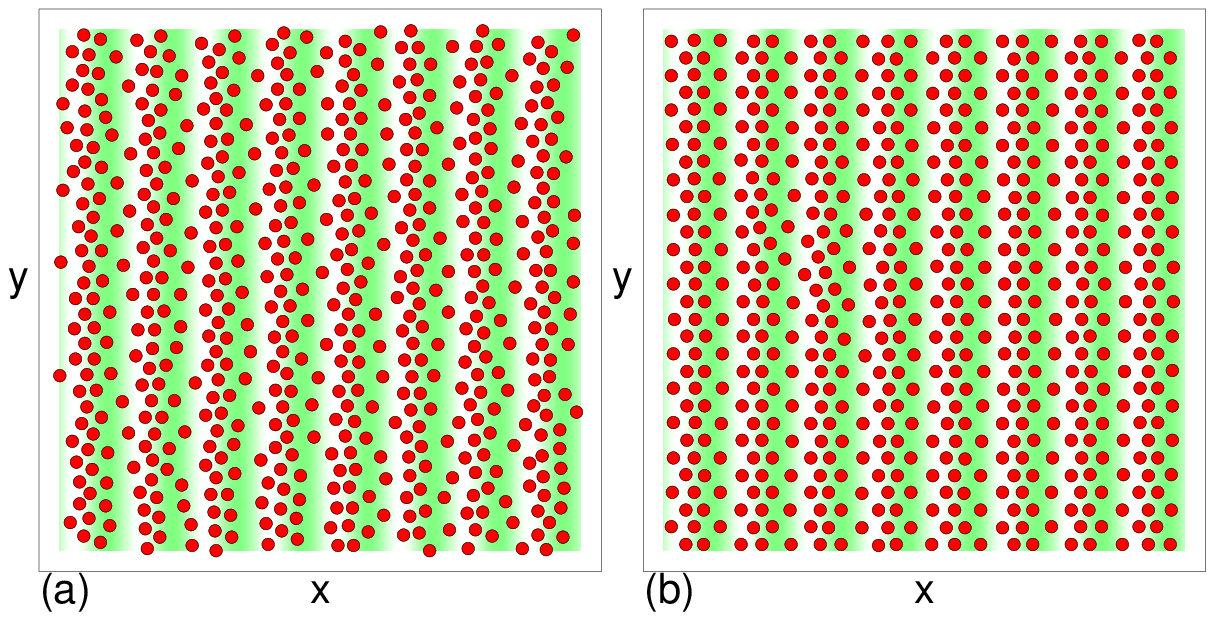}
\caption{Particle positions (red circles) and substrate potential
(green shading)
for the system from Fig.~\ref{fig:4} with
$\rho=0.44$, $F_p=1.0$, $N_p=8$, and $B=1.6$.  
(a) $F_{D} = 0.6$ in the partially disordered phase.
(b) $F_{D} = 1.5$ in the dynamically reordered
moving anisotropic crystal phase.
}
\label{fig:6}
\end{figure}

In Fig.~\ref{fig:4}, we show the $\langle V\rangle$ 
and $d\langle V\rangle/dF_{D}$ versus $F_{D}$
curves for the
system from Fig.~\ref{fig:2} in the anisotropic crystal state at
$B = 1.6$
and $B=0.2$.
When $B = 0.2$, the
particles undergo weak plastic depinning
from the anisotropic pinned crystal to a flowing disordered state,
and then transition
near $F_{D} = 0.4$
into a dynamically reordered moving crystal.
The corresponding $d\langle V\rangle/dF_D$ curve
contains only a single peak at the depinning transition.
At $B = 1.6$, the depinning is strongly plastic, and soliton-like flow
occurs, corresponding to the
nonlinear segment of the
$\langle V\rangle$ versus $F_{D}$ curve appearing below $F_D=0.4$.
As the drive increases,
a disordered flow regime appears, followed by
dynamical ordering into a moving crystal for $F_{D} > 0.65$.
In Fig.~\ref{fig:6}(a) we show the particle positions
in the disordered flowing state at
$B = 1.6$ and $F_{D} = 0.6$, and
in Fig.~\ref{fig:6}(b) we illustrate the dynamically reordered
moving anisotropic lattice at $F_D=1.5$. The lattice
displays a small density modulation from the substrate.

\begin{figure}
\includegraphics[width=\columnwidth]{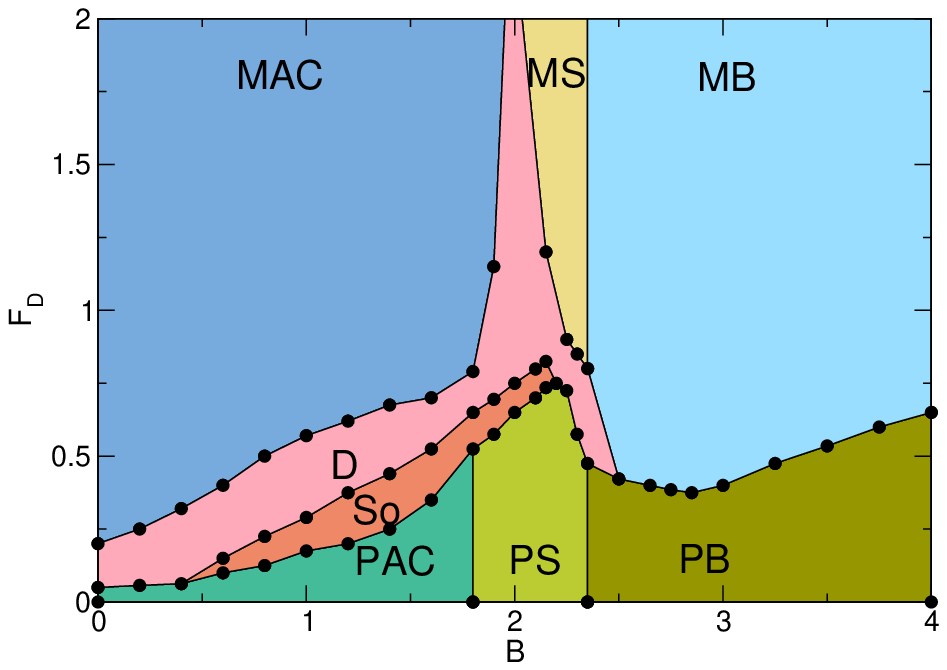}
\caption{Dynamic phase diagram as a function of $F_{D}$ vs $B$
constructed from the transport curves, pinned
structures, and moving structures for the system from
Fig.~\ref{fig:2} with $\rho = 0.44$, $N_{p} = 8$, and $F_{p} = 1.0$.
There are three pinned phases:  pinned anisotropic crystal (PAC),
pinned stripe (PS), and pinned bubble (PB).
The moving phases are
soliton flow (So), disordered flow (D),
moving anisotropic crystal (MAC),
moving stripe (MS), and moving bubble (MB).} 
\label{fig:7}
\end{figure}

From the features in the transport curves and the particle
arrangements, we can construct a dynamic phase
diagram of the different phases as a function of
$F_{D}$ versus $B$ for the sample with $\rho = 0.44$,
$N_{p} = 8$, and $F_{p}= 1.0$, as shown in Fig.~\ref{fig:7}.
The pinned states consist of 
the pinned anisotropic crystal (PAC), pinned stripe (PS),
and pinned bubble (PB) phases.
The PS
can depin into either a moving soliton (So) phase or
a disordered (D) plastic flow phase,
and it dynamically reorders into a moving stripe (MS) at high drives.
The PB phase shows
a region of plastic depinning near $B = 2.35$,
but for larger $B$ it elastically depins
directly into the moving bubble (MB) phase.
The PAC depins into the So phase
and then transitions into the D phase before
undergoing dynamic reordering into a moving anisotropic
crystal (MAC).
At small $B$, there is no soliton depinning and the PAC depins directly
into D flow.
An interesting feature is that
the drive needed to transition to the moving stripe
phase diverges near the MAC-MS boundary.
It is likely that the energy difference between the moving anisotropic
crystal and the moving stripe state is small near this boundary,
so the competition between the two phases causes a disordered flow to
emerge. Only at very high drives is it possible to resolve the energy
difference between the MAC and MS state, and escape from the disordered
flow at the MAC-MS boundary.

\section{Changing the Substrate Periodicity}

\begin{figure}
\includegraphics[width=\columnwidth]{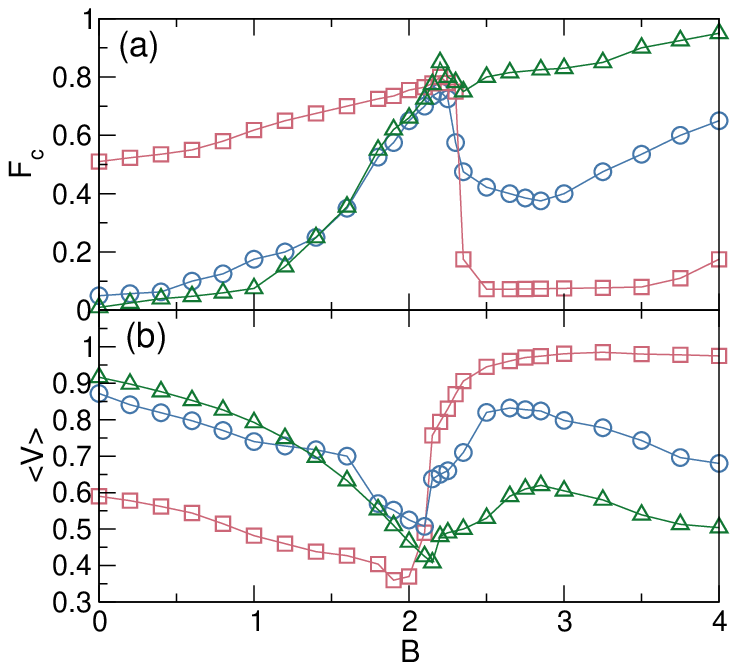}
\caption{(a) The depinning threshold $F_{c}$ vs
$B$ in samples with $\rho = 0.44$ and $F_{p}= 1.0$ at
$N_{p} = 4$ ($a_p=9.0$, green), $N_p=8$ ($a_p=4.5$, blue),
and $N_p=17$ ($a_p=2.1$, pink).
There is a maximum in $F_c$ in the stripe phase.
(b) The corresponding velocity $\langle V\rangle$ vs $B$
at $F_{D} = 1.0$ has a dip in the
stripe regime.
}
\label{fig:8}
\end{figure}

We next fix the particle density while
changing the number $N_p$ of substrate minima, which alters
the substrate lattice constant $a_p$.
In Fig.~\ref{fig:8}(a) we plot $F_{c}$ versus $B$
in samples with $\rho = 0.44$ and $F_{p}= 1.0$
at
$N_{p} = 4$ ($a_{p}= 9.0$),
$N_{p} = 8$ ($a_p = 4.5$),
and $N_{p} = 17$ ($a_{p} = 2.1$).
The $N_{p}= 8$ curve was already
highlighted in Fig.~\ref{fig:2}.
For $N_{p}= 4$, the anisotropic crystal phase
becomes even more weakly pinned since
the particles are able to fill all of the space;
however, the bubble phase is now strongly pinned since the bubbles
can easily fit within the substrate troughs. 
A small peak in $F_c$ appears near the stripe phase.
When $N_{p} = 17$, the anisotropic
crystal is strongly pinned, and the
pinned stripe configuration at $B=2.15$,
illustrated in Fig.~\ref{fig:1}(d), consists of rows that are a single
particle wide.
The bubble phase for $N_{p} = 17$
is weakly pinned because an individual
bubble has a radius that is much larger than
the pinning period of $a_p = 2.1$.
Figure~\ref{fig:8}(b) shows
the corresponding $\langle V\rangle$ versus $B$
curves at $F_{D} = 1.0$, where a local minimum
in $\langle V\rangle$ appears
in the stripe phase.
For $N_{p} = 17$, the bubbles
at higher values of $B$ slide at nearly
the expected free flow velocity of $\langle V\rangle = 1.0$.

\begin{figure}[h]
\includegraphics[width=\columnwidth]{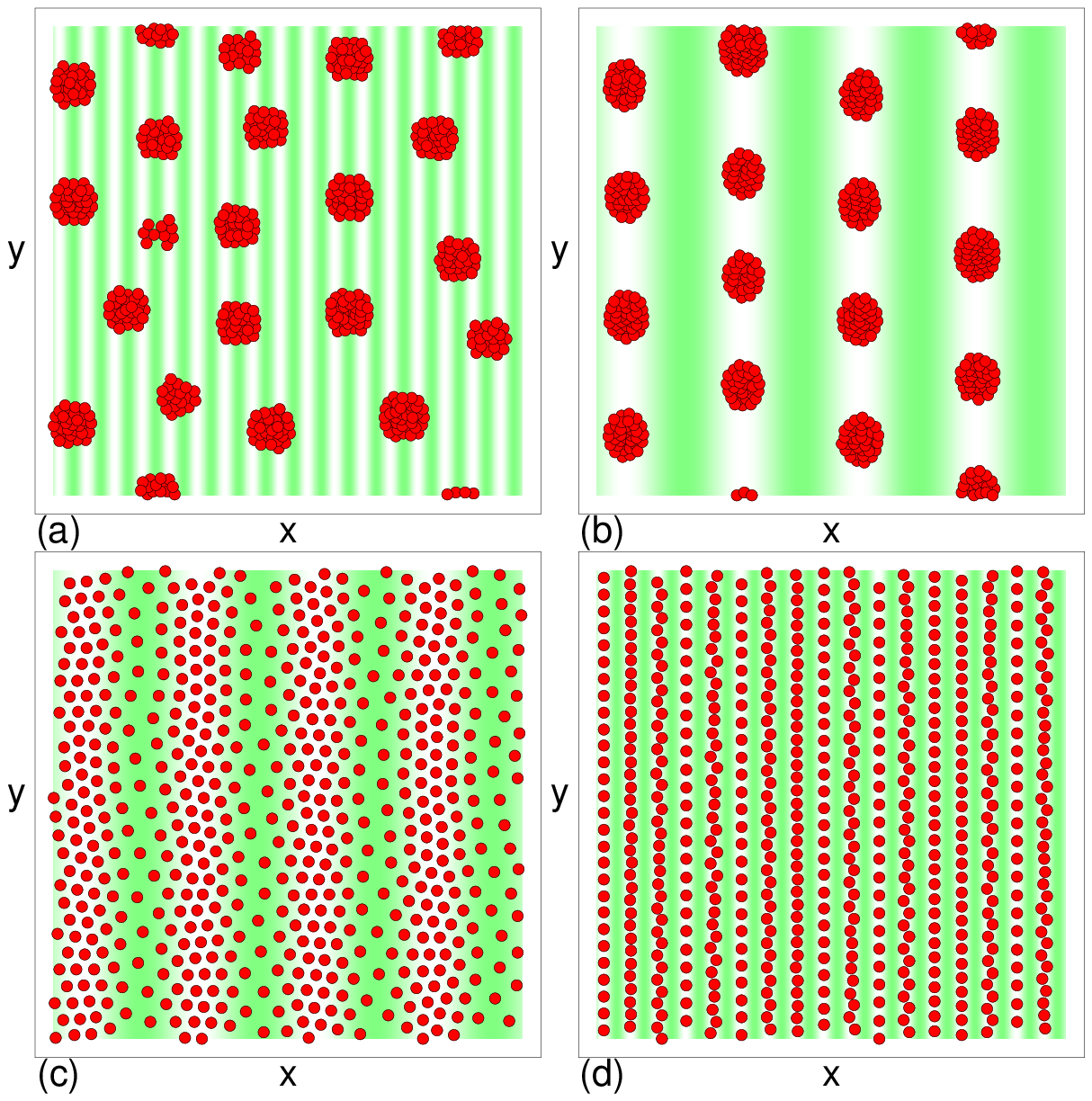}
\caption{Particle positions (red circles) and substrate potential
(green shading) for the system from Fig.~\ref{fig:8} with $\rho=0.44$ and
$F_p=1.0$ in the pinned state.  
(a) Weakly pinned bubbles at $B = 2.75$ and $N_{p} = 17$.
(b) Strongly pinned bubbles at $B = 2.75$ and $N_{p} = 4$.
(c) A weakly pinned anisotropic crystal at $B = 0.2$ and $N_{p} = 4.$
(d) A strongly pinned stripelike structure at $B = 0.2$ and $N_{p} = 17$.
}
\label{fig:9}
\end{figure}

In Fig.~\ref{fig:9}(a), we show the pinned particle configuration
for the bubble phase from Fig.~\ref{fig:8} at $B = 2.75$ and $N_{p} = 17$.
The bubble radius is twice as large as the substrate spacing, so the
bubbles can slide easily over the substrate.
Figure~\ref{fig:9}(b) shows the same system at $N_{p} = 4$,
where the bubbles easily
fit within the substrate troughs
and are strongly pinned.
At $B = 0.2$ and $N_{p}= 4$
in Fig.~\ref{fig:9}(c), 
the system forms
a density modulated crystal that is weakly pinned since
some of the particles are located near maxima of the substrate
potential.
In Fig.~\ref{fig:9}(d), the particle configuration at $B = 0.2$ and $N_{p}= 17$
adopts a stripe-like pattern that is almost the
same as the configuration
found in Fig.~\ref{fig:2} for $B = 2.15$ and the same substrate spacing.
Fig.~\ref{fig:8}(a) shows that
the depinning threshold
remains nearly constant
for the anisotropic crystal and
stripe states over the range $0 < B < 2.25$,
and only drops
in the bubble phase
when $B \geq 2.25$.

\begin{figure}
\includegraphics[width=\columnwidth]{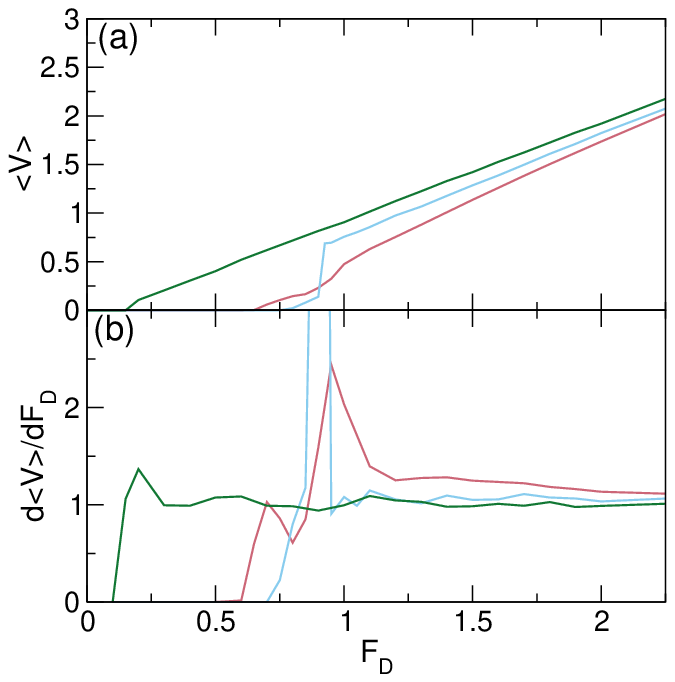}
\caption{(a) $\langle V\rangle$ vs $F_D$ 
for a system with $N_{p} = 17$, $\rho = 0.44$, and $F_{p} = 1.0$
at $B = 1.0$ (red), $2.15$ (blue), and $2.75$ (green).
(b) The corresponding $d\langle V\rangle/dF_{D}$ vs $F_{D}$ curves.} 
\label{fig:10}
\end{figure}

\begin{figure}
\includegraphics[width=\columnwidth]{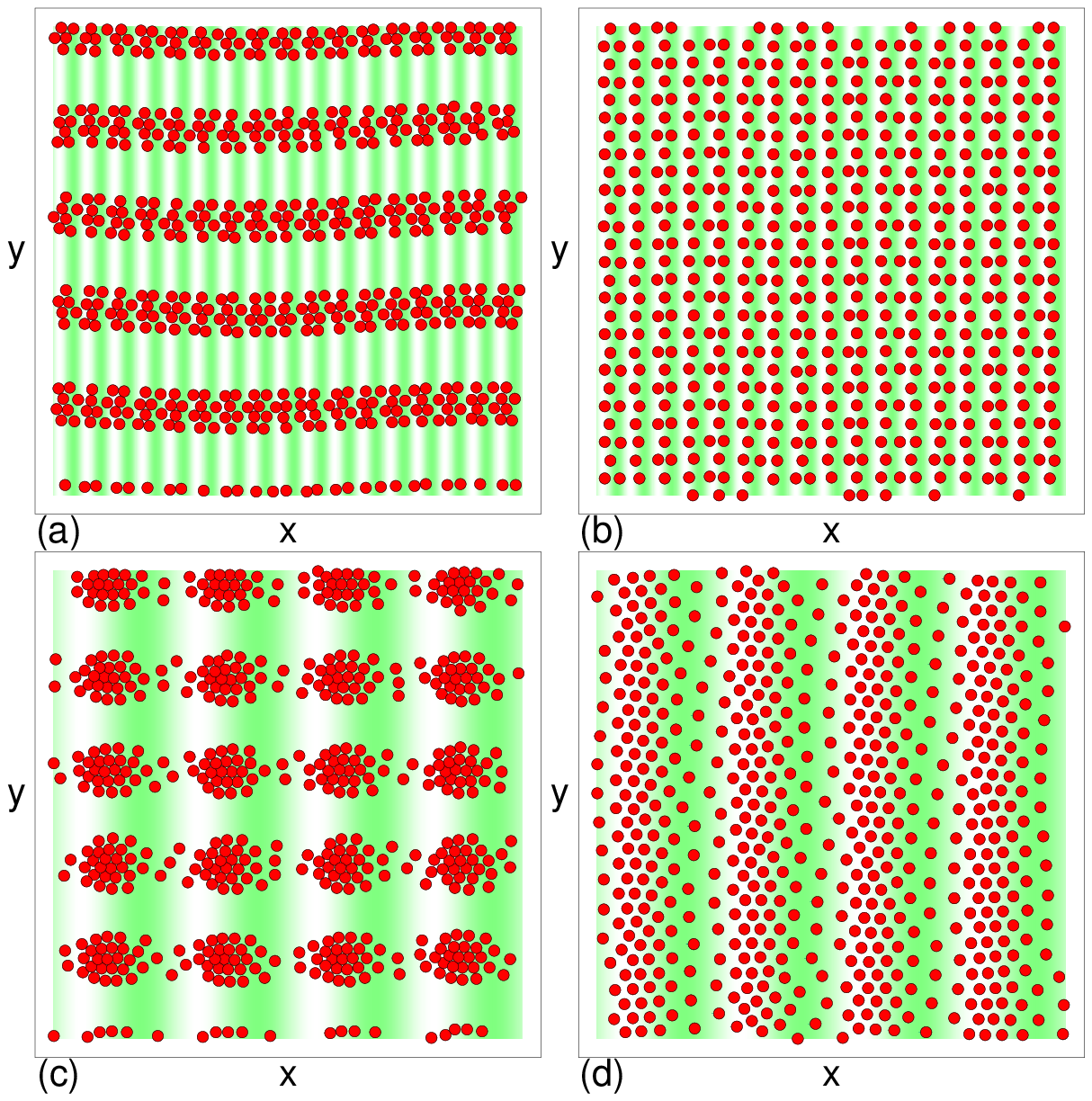}
\caption{Particle positions (red circles) and substrate potential (green
shading) for the system from Fig.~\ref{fig:10} with $\rho=0.44$,
$F_p=1.0$, and $F_D=1.2$ for varied $B$ and $N_p$.  
(a) The moving stripe state at $N_p=17$ and $B=2.15$.
(b) The moving crystal phase at $N_p=17$ and $B = 1.0$.
(c) The moving stripe state at $N_{p} = 4$ and $B = 2.15$, 
where the stripes are more bubble-like.
(d) The moving modulated solid at $N_{p}  = 4$ and $B = 1.0$.
} 
\label{fig:11} 
\end{figure}

For $N_{p}= 17$, we observe dynamical phases similar to those
described above for the $N_{p}= 8$ system.
There are, however,
some differences in the transport curves, 
as shown in Fig.~\ref{fig:10}(a) where we plot $\langle V\rangle$
versus $F_D$ at $B=1.0$, 2.15, and 2.75.
When $B = 2.15$,
there is a two-step depinning process from
a soliton like flow to a moving stripe state. The second depinning
transition is more discontinuous than the first,
resulting in the appearance in Fig.~\ref{fig:10}(b) of
a strong peak in $d\langle V\rangle/dF_D$ versus $F_D$ 
at the onset of the stripe phase.
For $B = 1.0$, there is also
a two-step depinning process that produces a double peak in 
$d\langle V\rangle/dF_{D}$, but neither of the peaks are as sharp
as the peak found in the $B = 2.15$ sample.
At $B = 2.75$, there is a single peak in
$d\langle V\rangle/dF_{D}$, and the pinned bubble phase
depins elastically to a moving bubble phase.
Another interesting feature
is that even though the depinning threshold is largest for $B = 2.15$,
the $B=1.0$ and $B=2.15$ velocity-force curves cross
at higher drives so that, within the moving stripe regime,
the velocity is higher for $B=2.15$ than for $B=1.0$,
indicating that for this value of $N_p$
the stripes can flow with less resistance
than the moving anisotropic crystal.
In Fig.~\ref{fig:11}(a), we illustrate the moving stripe state
for the system from Fig.~\ref{fig:10}
with $N_p=17$ at $B=2.15$ and $F_{D} = 1.2$,
where the stripes are aligned with the driving
direction and are four particles wide.
Figure~\ref{fig:11}(b) shows the moving crystal phase
at $B = 1.0$ and $F_{D} = 1.2$. 
When $N_{p} = 4$, a similar set of phases
occurs but the features in the transport curves
are not as sharp.
In Fig.~\ref{fig:11}(c) we show the $N_p=4$ stripe phase
at $B  = 2.15$ and $F_{D}= 1.2$,
where the stripes are more bubble-like.
Figure~\ref{fig:11}(d) shows the modulated moving solid at
$B = 1.0$ and $F_{D} = 1.2$ in the $N_p=4$ system.

\section{Effect of Substrate Strength}

\begin{figure}
\includegraphics[width=\columnwidth]{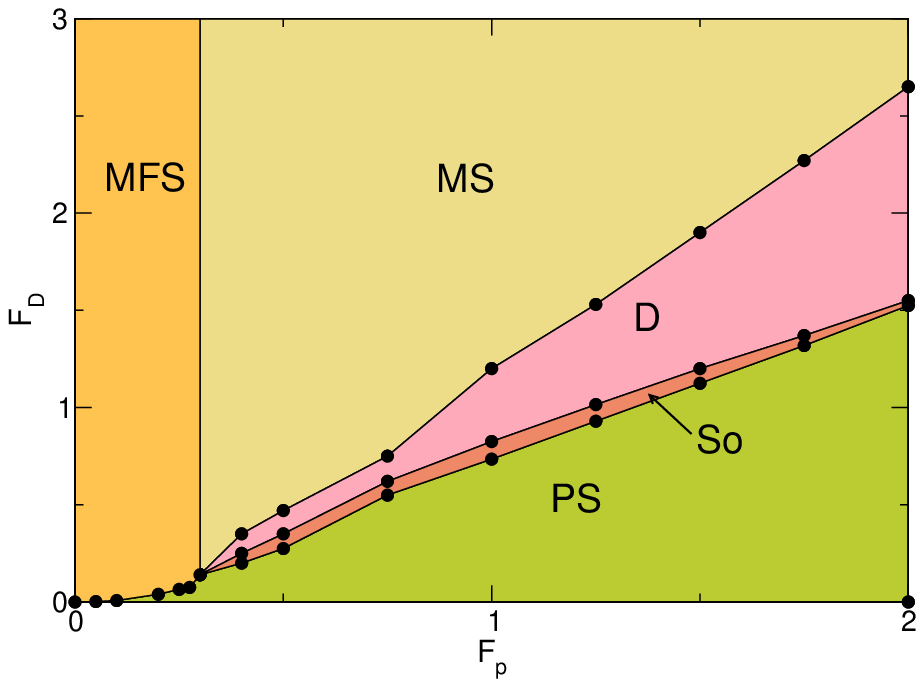}
\caption{Dynamic phase diagram as a function of
$F_{D}$ vs $F_{p}$ for
a stripe forming system at $\rho=0.44$, $B = 2.15$,
and $N_{p} = 8$ showing the pinned stripe (PS), soliton flow (So),
disordered flow (D), moving stripe (MS), and moving floating solid (MFS).
For $F_{p} \leq 0.35$, the PS depins elastically to a MFS. For
$F_{p} > 0.35$ the PS depins plastically to the So state,
then transitions into a D phase and finally a MS state.
}  
\label{fig:12}
\end{figure}

In general, changing the substrate strength does not modify which
dynamic phases are present, but it shifts the boundaries between
the phases.
As a function of increasing pinning strength,
for the bubble phases we find
a transition from elastic to plastic depinning
at a critical pinning force that is accompanied
by a noticeable increase in $F_{c}$ \cite{Reichhardt24}. 
For the anisotropic crystal, we observe a similar step up in $F_c$
at an elastic to plastic depinning transition, 
but it is not as pronounced as in the bubble phases.
In Fig.~\ref{fig:12} we plot a dynamic phase diagram as
a function of $F_D$ versus $F_p$ for the stripe state at 
$B = 2.15$, $\rho = 0.44$, and $N_{p} = 8$.
For $F_{p} \leq 0.35$, the stripes depin elastically and do not
rotate to align with the driving direction but instead enter what
we term a moving floating solid (MFS) state.
For $F_{p} >0.35$, the system
depins plastically.
There is a small jump up in the depinning threshold at the elastic-to-plastic
depinning transition,
and $F_{c}$ increases linearly with increasing $F_{p}$
in the plastic depinning regime. As the drive increases above the plastic
depinning transition, the system first passes through a soliton flow phase
and then into a disordered flow state before reaching a moving stripe state
in which the stripes are aligned with the driving direction.

\begin{figure}
\includegraphics[width=\columnwidth]{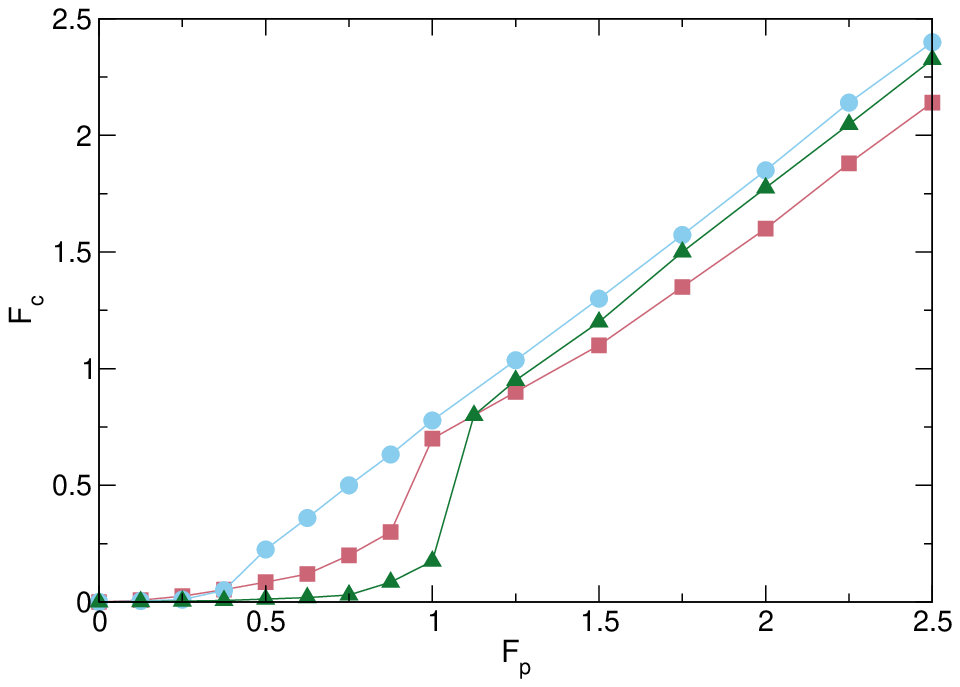}
\caption{$F_{c}$ vs $F_{p}$ for a system with $\rho = 0.44$ and
$N_{p} = 17$ at $B = 1.6$ (anisotropic crystal, red squares),
$B = 2.15$ (stripe, blue circles) and $B = 2.35$ (bubble, green triangles).
}  
\label{fig:13}
\end{figure}

When $N_{p} = 17$, the transition from elastic to plastic depinning
as a function of increasing $F_{p}$
is much sharper than in the $N_p=8$ system.
In Fig.~\ref{fig:13} we plot $F_{c}$ versus $F_{p}$ for a system
with $\rho = 0.44$ and
$N_{p} = 17$ at
$B = 1.6$ in the anisotropic crystal state,
$B = 2.15$ in the stripe state, and $B=2.35$ in the bubble state.
The depinning threshold is largest for the stripe system
at all $F_{p}$, and the stripe state undergoes an elastic 
to plastic depinning transition 
near $F_{p} = 0.35$.
For $B = 2.35$, the bubbles depin elastically
up to $F_{p} = 1.0$, and then
a large increase in the depinning threshold occurs at the transition to
plastic depinning.
For $B  = 1.6$, the transition from elastic to plastic
depinning appears at $F_p = 0.9$ and is accompanied by a large increase
in $F_c$.
Within the plastic depinning regime,
the depinning threshold increases linearly with
increasing $F_{p}$ as found for
other interacting particle systems \cite{Reichhardt17},
while in the elastic depinning regime, $F_c$ is a nonlinear function of $F_p$.
Our results are not accurate enough to give the exact fitting
function, but systems that depin elastically often have a depinning
threshold that increases quadratically with increasing
pinning force \cite{Reichhardt17}.

\begin{figure}
\includegraphics[width=\columnwidth]{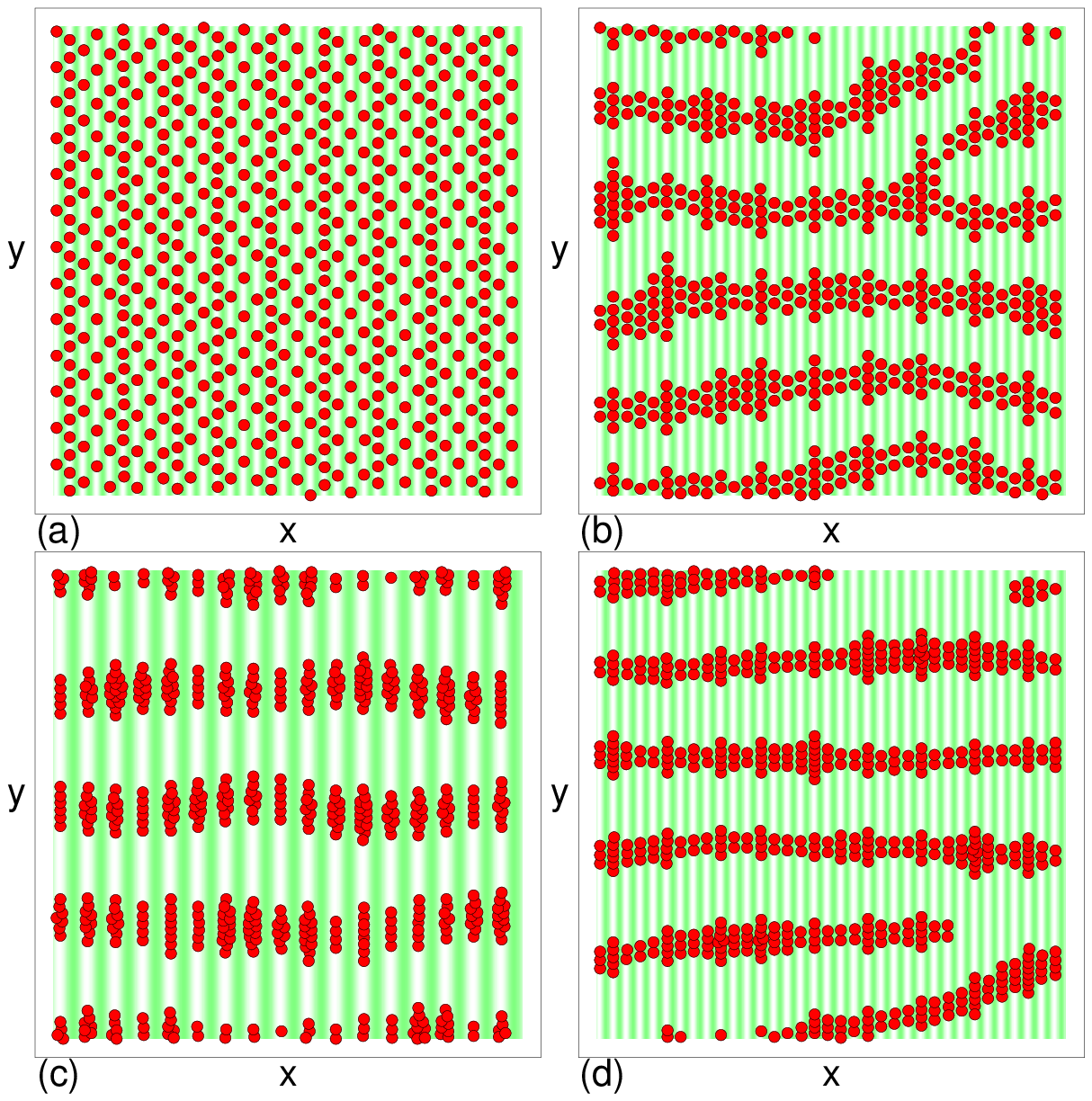}
\caption{Particle positions (red circles) and substrate potential (green shading)
showing pinned configurations at $F_D=0$ and $\rho=0.44$. 
(a) A distorted crystal at $N_p = 35$, $F_{p} = 2.5$, and $B = 1.6$.
(b) A pinned stripe aligned with the $x$ direction at $N_p = 35$, $F_{p} = 2.5$,
and $B = 2.15$.
(c) A pinned modulated stripe aligned with the $x$ direction
at $N_{p} = 17$, $F_{p} = 3.0$ and $B = 2.75$.
(d) A stripe-like structure at $N_{p} = 35$, $F_{p} = 1.5$, and $B = 2.75$.
}
\label{fig:14} 
\end{figure}

\begin{figure}
\includegraphics[width=\columnwidth]{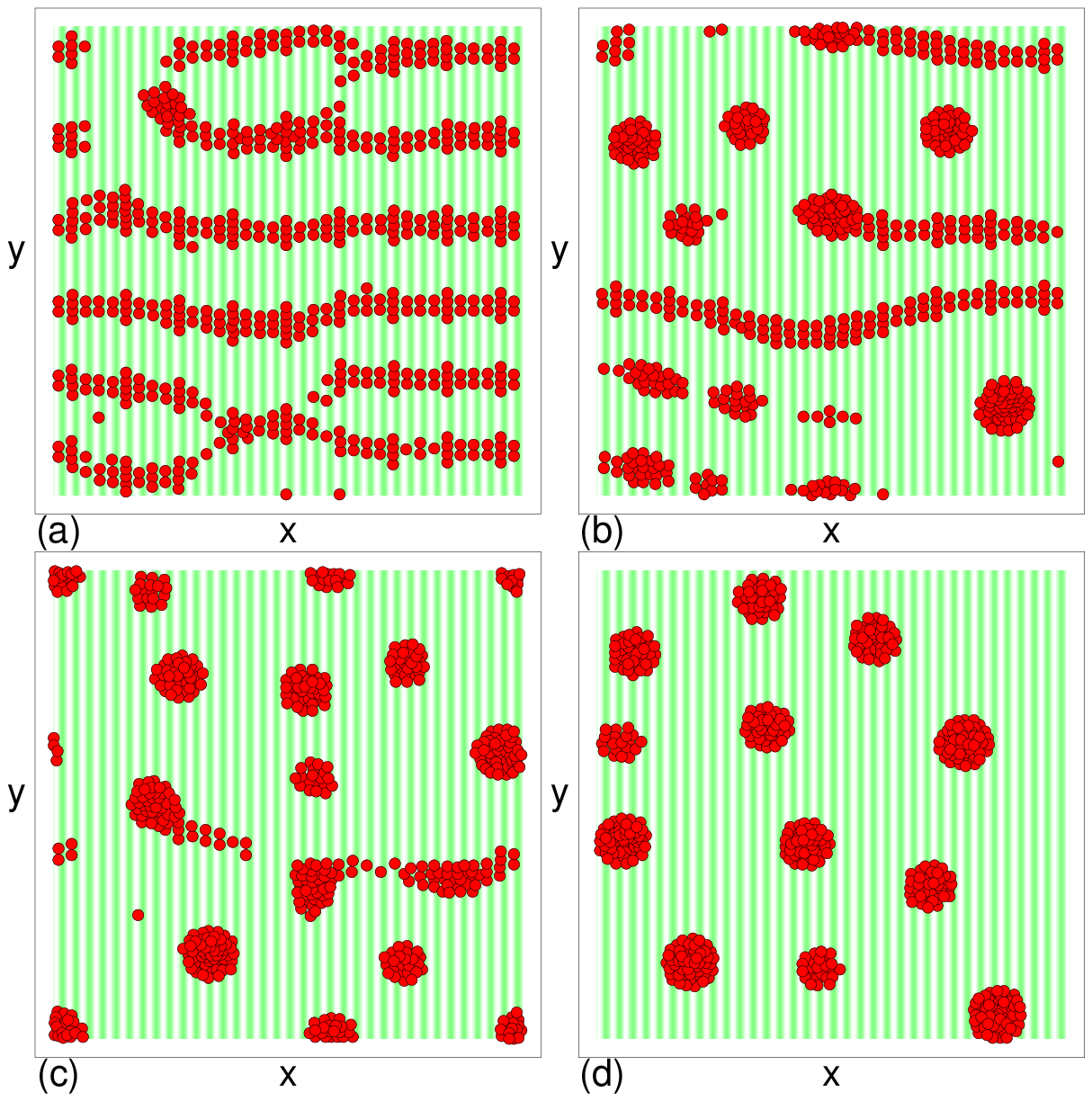}
\caption{Particle positions (red circles) and substrate potential (green
shading) showing the 
time evolution of the pinned stripe into a moving bubble for
$F_{p}= 3.0$,  $B = 2.75$, $N_{p} = 35$, $\rho=0.44$, and $F_{D} = 2.0$.
(a) At early time, nascent bubbles begin translating along the stripe
structures and collecting additional particles.
(b) At early intermediate time, more of the pinned stripe particles become
incorporated into the moving bubbles.
(c) At late intermediate time, very few of the particles are still
in pinned stripe structures.
(d) At late time, all of the particles have joined
moving bubbles.
}
\label{fig:15}
\end{figure}

When $B > 2.0$,
another effect we observe is that for stronger substrates
and higher $N_{p}$,
the system forms pinned stripes or modulated stripes
that are perpendicular to the substrate troughs.
In Fig.~\ref{fig:14}(a),
we illustrate the pinned configuration for
a system with $N_p = 35$, $F_{p} = 2.5$,
and $B = 1.6$, where a distorted crystal state appears.
Figure~\ref{fig:14}(b) shows the same system at
$B = 2.15$,
where a pinned stripe appears that is aligned in the $x$ direction,
perpendicular to the substrate troughs.
At $N_{p} = 17$, $F_{p} = 3.0$, and $B = 2.75$ in Fig.~\ref{fig:14}(c),
we find a more discontinuous modulated stripe-like pattern aligned with
the $x$ direction, while
a stripe-like pinned pattern appears
for $N_{p} = 35$, $F_{p} = 1.5$, and $B = 2.75$ in
Fig.~\ref{fig:14}(d).
For either value of $N_p$,
when $B = 2.15$, the system can still depin plastically via
solitons that run along the stripe, but as the drive increases,
the entire stripe structure depins and remains
aligned in the direction of the drive. 
For $B = 2.75$, the moving state forms bubbles,
and depinning occurs via the formation of a bubble along the modulated
stripe that picks up particles as it travels along the stripe as if
the stripe is providing a track for motion.
In Fig.~\ref{fig:15} we illustrate the time evolution of the depinning
of the stripe state into a moving bubble state for a sample with
$F_{p}= 3.0$,  $B = 2.75$, $N_{p} = 35$, $\rho=0.44$,  and $F_{D} = 2.0$.
At early times in Fig.~\ref{fig:15}(a),
the depinning occurs via the formation of bubbles
that move along the stripe structures.
The number of bubbles grows at later times, as shown in
Fig.~\ref{fig:15}(b,c), and at long
times all of the particles are
contained by moving bubbles as in Fig.~\ref{fig:15}(d).

\section{Varied Filling}

\begin{figure}
\includegraphics[width=\columnwidth]{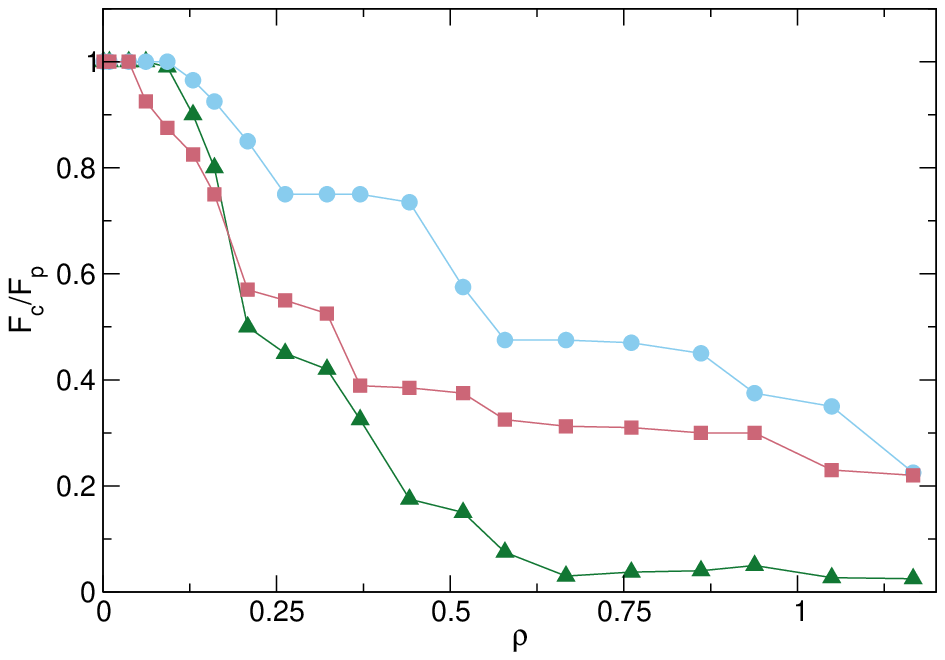}
\caption{The critical depinning force $F_{c}/F_{p}$ vs
particle density $\rho$ for a system
with $F_{p} = 1.0$ and $N_p=8$
at $B = 2.15$ (stripes, blue circles),
$B = 2.75$ (bubbles, red squares), and $B = 1.0$
(anisotropic crystal, green triangles).
} 
\label{fig:16} 
\end{figure}

\begin{figure}
\includegraphics[width=\columnwidth]{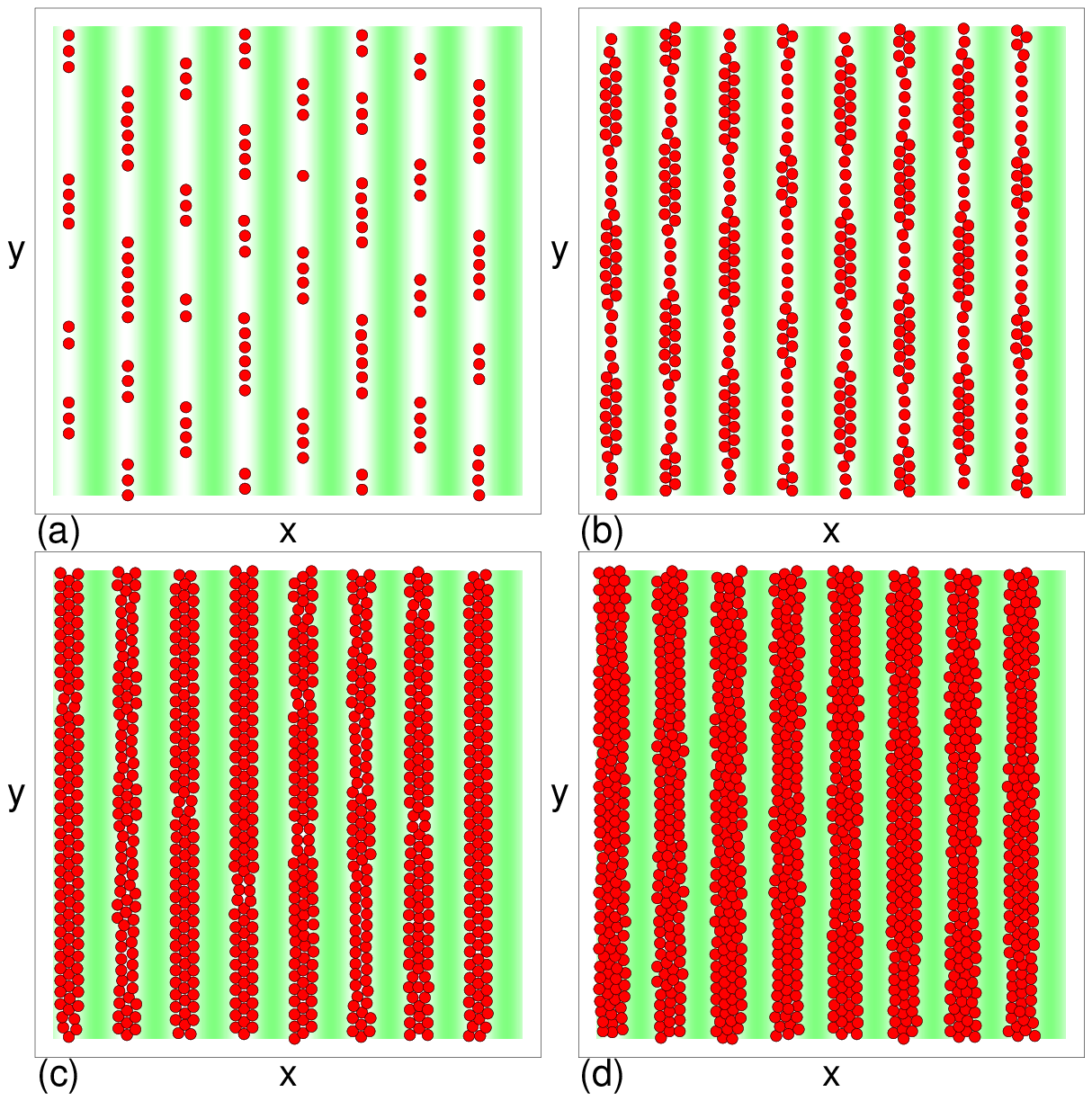}
\caption{Particle positions (red circles) and substrate potential (green
shading) showing the pinned particle configurations
for the system from Fig.~\ref{fig:16} with $F_p=1.0$, $N_p=8$,
and $B=2.15$ in the stripe state at
(a)  $\rho = 0.093$, (b) $\rho = 0.262$,
(c) $\rho = 0.67$, and (d) $\rho = 0.938$.
}
\label{fig:17}
\end{figure}

We next consider the effect of holding $B$, the substrate strength, and
the substrate lattice constant fixed while 
varying the particle density $\rho$.
In Fig.~\ref{fig:16}, we plot
$F_{c}/F_{p}$ versus $\rho$ for
samples with $F_{p} = 1.0$ and $N_{p} = 8.0$ at
$B = 2.15$ in the stripe state, $B = 2.75$ in the bubble state,
and $B = 1.0$ in the anisotropic crystal state.
The stripe state has the highest depinning threshold across the
entire range of $\rho$, and shows some plateaus with
$F_{c}/F_{p} = 1.0$ for $\rho < 0.1$,
$F_c/F_p \approx 0.75$ for $0.25 < \rho < 0.5$,
$F_c/F_p = 0.47$ for $0.55 < \rho < 0.85$, and a decrease
in $F_c/F_p$ at higher values of $\rho$.
These plateaus correspond to stripes that are composed of different numbers
of particles per row.
In Fig.~\ref{fig:17}(a) we plot the pinned particle configurations
for $B = 2.15$ at $\rho = 0.093$, where $F_{c}/F_{p} = 1.0$.
Here, the system forms a stripelike bubble state where each bubble is
only a single particle wide and the disordered bubbles are arranged
in a rough lattice configuration.
Figure~\ref{fig:17}(b) shows the
pinned configuration at $\rho = 0.262$ on the second
plateau in $F_c/F_p$,
where a continuous stripe structure appears in which some
regions of the stripe are two particles wide.
It is the regions of greater width
that depin first.
In general, the second plateau in $F_c/F_p$
for the $B = 2.15$ system corresponds to values of $\rho$ for which
portions of the stripe are two particles wide.
On the third plateau in $F_c/F_p$, as shown in
Fig.~\ref{fig:17}(c) at $\rho = 0.67$,
portions of the stripes are three particles wide.
Figure~\ref{fig:17}(d) shows the
pinned stripe configuration at $\rho = 0.938$, where the stripes now have
a width of four particles.
In general, we expect that there should be a series of plateaus
whenever $N$ rows of particles can fit inside one of the substrate troughs.

\begin{figure}
\includegraphics[width=\columnwidth]{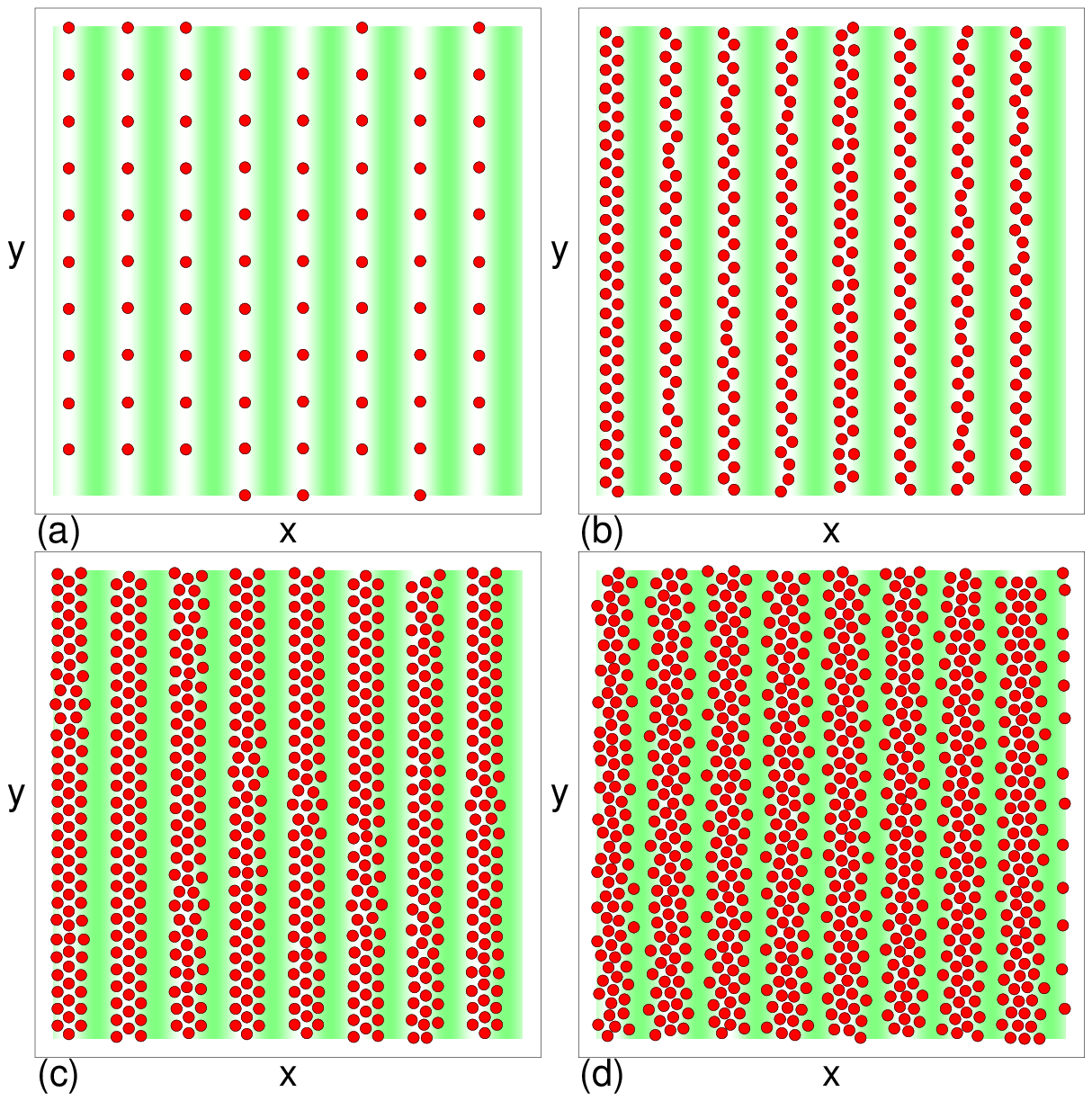}
\caption{Particle positions (red circles) and substrate potential
(green shading) showing the pinned
anisotropic crystal system from Fig.~\ref{fig:16} with
$F_p=1.0$, $N_p=8$, and $B=1.0$ at  
(a)  $\rho = 0.093$, (b) $\rho = 0.262$,
(c) $\rho = 0.518$, and (d) $\rho = 0.67$.
}
\label{fig:18}
\end{figure}

In Fig.~\ref{fig:18}(a), we show the pinned particle configurations for
the system from Fig.~\ref{fig:16} in the $B=1.0$ anisotropic
crystal state at $\rho = 0.093$, where the particles form a
rectangular array. Here, triangular ordering is suppressed by the
attractive interaction term; if the particle interactions were purely
repulsive, the particles would try to move as far away from each other
as possible and would adopt a triangular configuration, but the attractive
term causes square or rectangular configurations to be favored.
Figure~\ref{fig:18}(b) shows the same system
at $\rho = 0.262$, where stripes containing two rows of particles
have formed. The pattern is more zig-zag in nature compared to the
$B=2.15$ system,
and there are no regions where the stripes are strictly 1D-like.
At $\rho=0.518$ in Fig.~\ref{fig:18}(c),
there are now three rows of
particles in each substrate minimum.
For the even higher density of
$\rho = 0.67$, there is not enough space in the substrate minima to
accommodate all of the particles, so the system becomes partially
disordered and the depinning threshold drops considerably.

\begin{figure}
\includegraphics[width=\columnwidth]{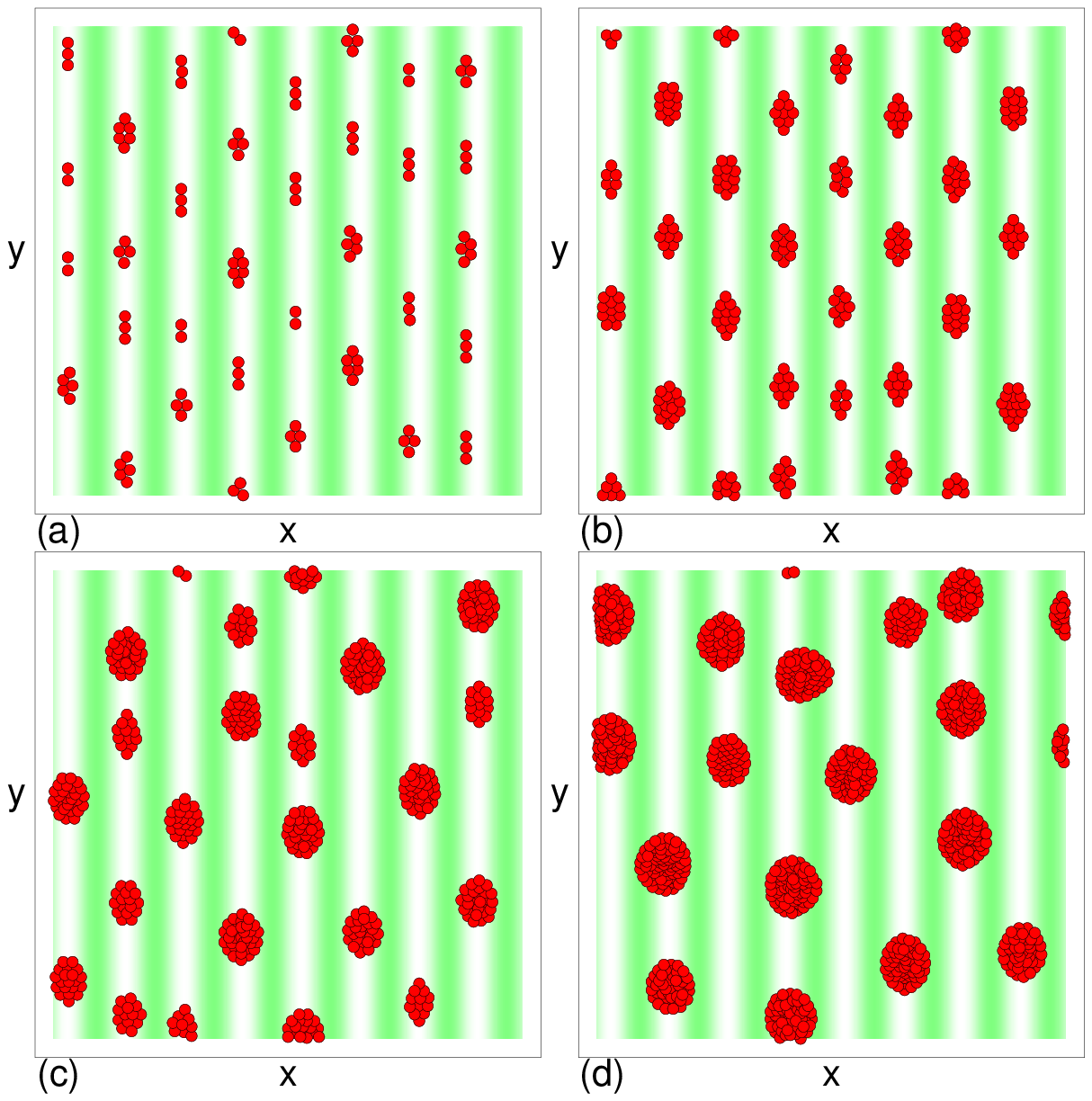}
\caption{Particle positions (red circles) and substrate potential
(green shading) showing the pinned particle configurations
for the $B=2.75$ bubble forming system
from  Fig.~\ref{fig:16} with $F_{p} = 1.0$ and
$N_{p} = 8$. (a) $\rho = 0.093$. (b) $\rho = 0.208$.
(c) $\rho = 0.37$. (d) $\rho = 0.76$.
}
\label{fig:19}
\end{figure}

Figure~\ref{fig:19}(a) shows the particle configurations
in the pinned bubble state for the system from Fig.~\ref{fig:16} 
at $B = 2.75$ and $\rho=0.093$, where a series of small bubbles appear.
For this filling, the depinning threshold is smaller than in
the stripe and anisotropic crystal states, where the particle arrangements
were strictly 1D.
At $\rho=0.208$ in Fig.~\ref{fig:19}(b), 
the depinning threshold has dropped onto the next plateau,
and the bubbles have a width of three particles.
On the next plateau of the depinning threshold, illustrated
in Fig.~\ref{fig:19}(c) at
$\rho = 0.37$, the bubbles are much larger.
Finally, for $\rho=0.76$ in Fig.~\ref{fig:19}(d),
there are even larger bubbles.

\begin{figure}
\includegraphics[width=\columnwidth]{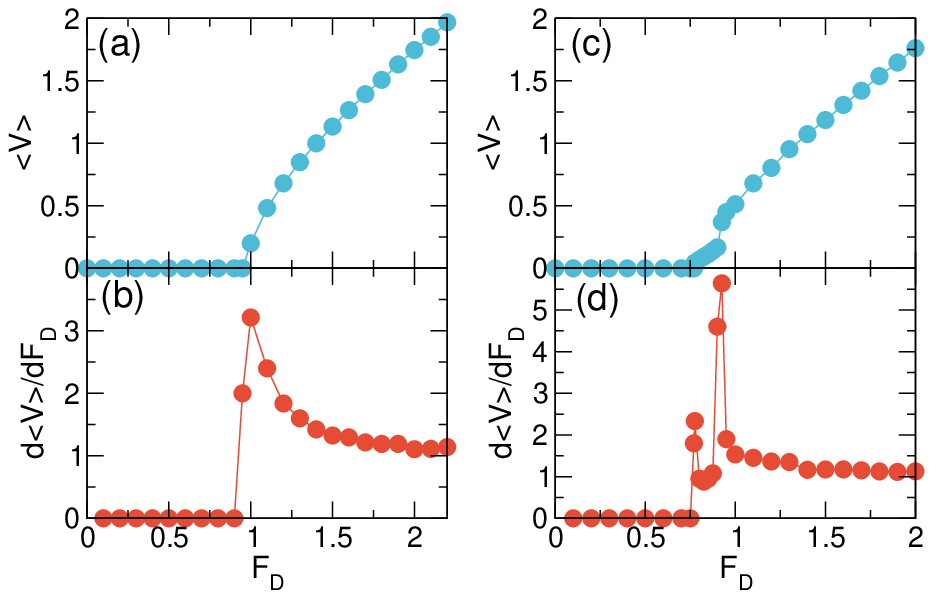}
\caption{(a,c) $\langle V\rangle$ vs $F_{D}$ and
(b,d) $d\langle V\rangle/dF_{D}$ vs $F_{D}$
for the system from Fig.~\ref{fig:16} with $F_p=1.0$, $N_p=8$,
and $B = 2.15$.
(a,b) $\rho = 0.093$.
(c,d) $\rho = 0.26$, where there is a two step depinning process.
}
\label{fig:20}
\end{figure}

\begin{figure}
\includegraphics[width=\columnwidth]{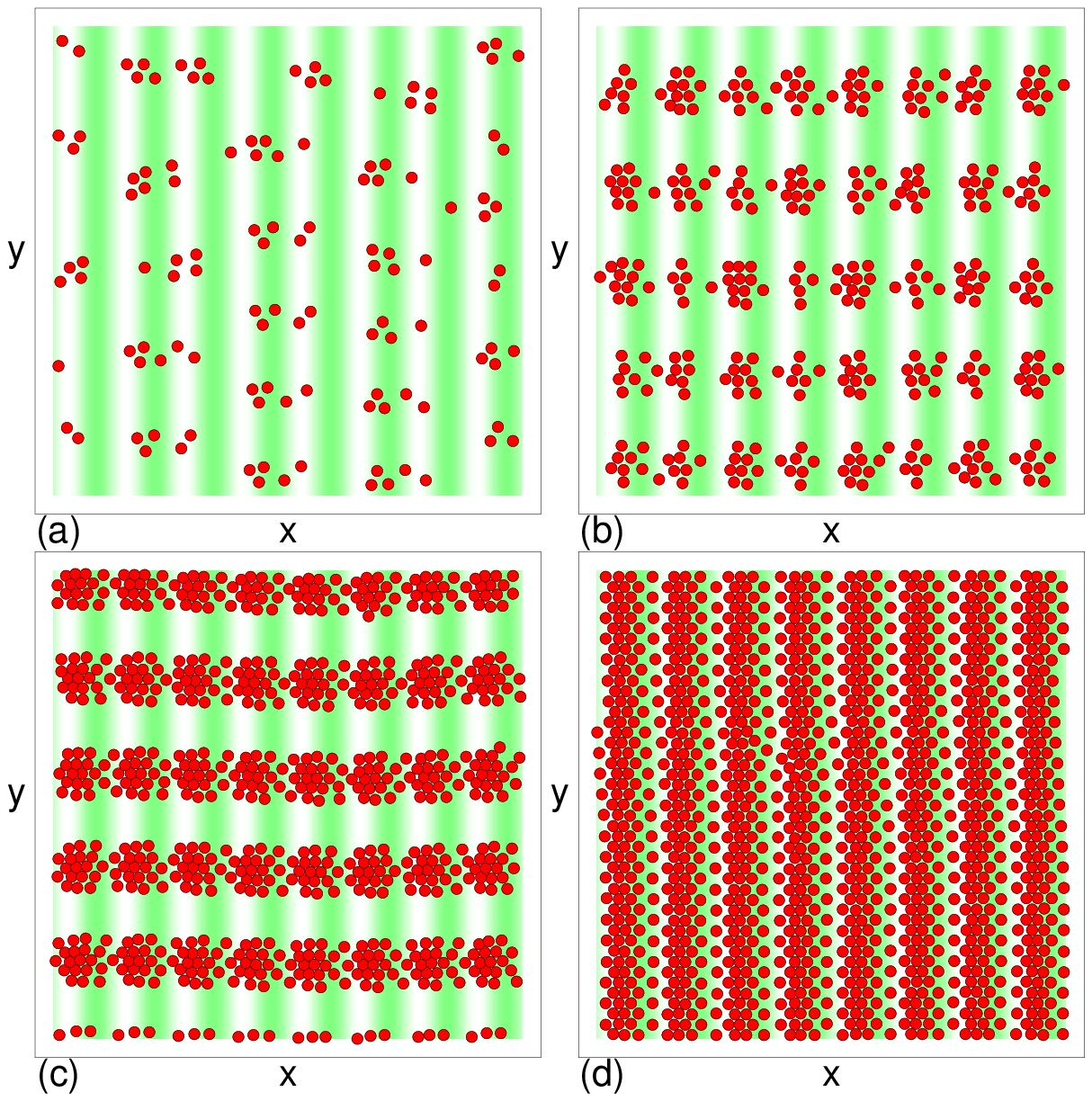}
\caption{Particle positions (red circles) and substrate potential (green
shading) in a series of moving states
at $F_D=1.5$ in samples with 
$F_p=1.0$, $N_p=8$, and $B=2.15$.
(a) The moving clump phase for the system
from Fig.~\ref{fig:20}(a,b)
with $\rho=0.093$.
(b) The moving stripe phase
for the system in Fig.~\ref{fig:20}(c,d)
with $\rho=0.26$.
(c) The moving
stripe state at $\rho = 0.578$ for the system in
Fig.~\ref{fig:22}(a,b) with $\rho=0.578$.
(d) The moving modulated solid
at $\rho = 0.938$
for the system in Fig.~\ref{fig:22}(c,d).
}
\label{fig:21}
\end{figure}

In Fig.~\ref{fig:20}(a,b) we plot $\langle V\rangle$ and
$d\langle V\rangle/dF_{D}$  versus $F_{D}$ for the stripe system
from Fig.~\ref{fig:16} 
at $B = 2.15$ and
$\rho = 0.093$.
The depinning has the character of a single-particle process, and
the system transitions
from a pinned 1D bubble state to a moving dilute bubble phase.
Here the depinning threshold falls slightly below $F_{D}/F_{p} = 1.0$.
At these low densities, there are not enough particles present to
permit a stripe phase to form,
but there are still some bubble-like features in the moving state. 
Fig.~\ref{fig:21}(a) illustrates the moving clump phase
for the system in Fig.~\ref{fig:20}(a) at $F_{D} = 1.5$.
The plots of $\langle V\rangle$ and $d\langle V\rangle/dF_D$ versus
$F_D$ in Fig.~\ref{fig:20}(b,c)
for a sample with $B = 2.15$ and
$\rho = 0.26$ indicate that there is
a two-step depinning process accompanied by
a double peak feature in the differential velocity-force curve.
The system first depins into a soliton motion state in which
the solitons translate through
the regions containing two rows of particles.
At higher drives, all of the particles depin and
the system forms a moving stripe phase, as shown
in Fig.~\ref{fig:21}(b) at $F_{D} = 1.5$.

\begin{figure}
\includegraphics[width=\columnwidth]{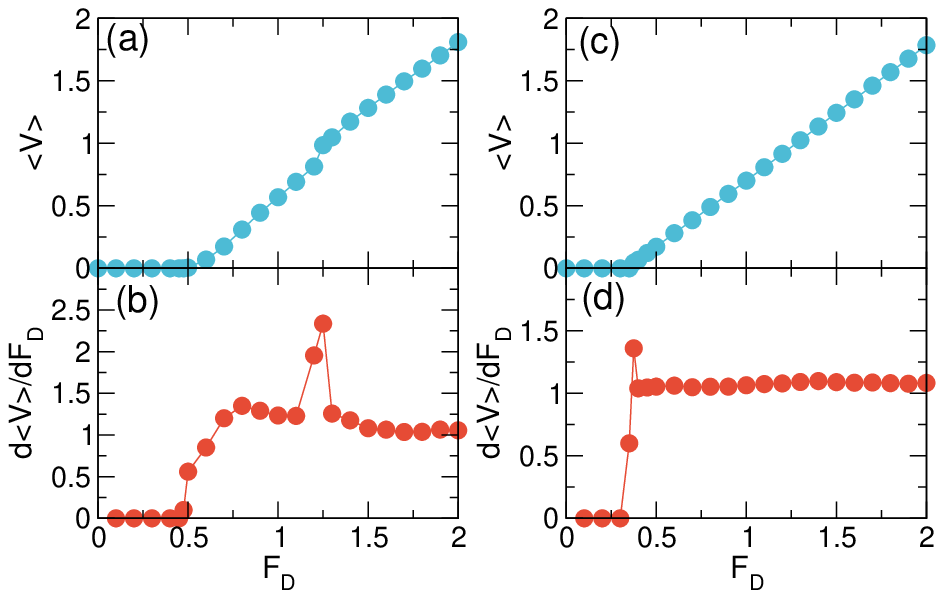}
\caption{(a,b) $\langle V\rangle$ vs $F_{D}$ and
(c,d) $d\langle V\rangle/dF_{D}$ vs $F_{D}$
for the system from
Fig.~\ref{fig:16} with
$F_p=1.0$, $N_p=8$, and $B = 2.15$.
(a,b) $\rho = 0.578$.
(c,d) $\rho = 0.938$, where there is a single elastic depinning process.
}
\label{fig:22}
\end{figure}

In Fig.~\ref{fig:22}(a,b) we plot $\langle V\rangle$ and
$d\langle V\rangle/dF_{D}$ versus $F_{D}$ for 
the system from Fig.~\ref{fig:20} with $B = 2.15$ at $\rho = 0.578$.
Here we observe a soliton depinning process, a second depinning
transition into a disordered flow state, and a dynamical reordering
transition into a moving stripe state,
visible as a peak in the differential mobility
near $F_D=1.25$.
Figure~\ref{fig:21}(c) shows the particle configuration
in the moving stripe state at
$F_D=1.5$ for $\rho = 0.578$, where each stripe has a 
width of five particles. 
The plots of $\langle V\rangle$ and $d\langle V\rangle/dF_D$
in Fig.~\ref{fig:22}(c,d)
for the same sample at
$\rho=0.938$ exhibit a single peak in the differential mobility produced
when
the system depins elastically from a modulated solid to
a moving modulated solid.
Here the stripes do not reorient into the direction of driving and
remain parallel to the substrate troughs.
Figure~\ref{fig:21}(d) illustrates the moving modulated solid  state
for $F_D=1.5$ at $\rho = 0.938$.

\begin{figure}
\includegraphics[width=\columnwidth]{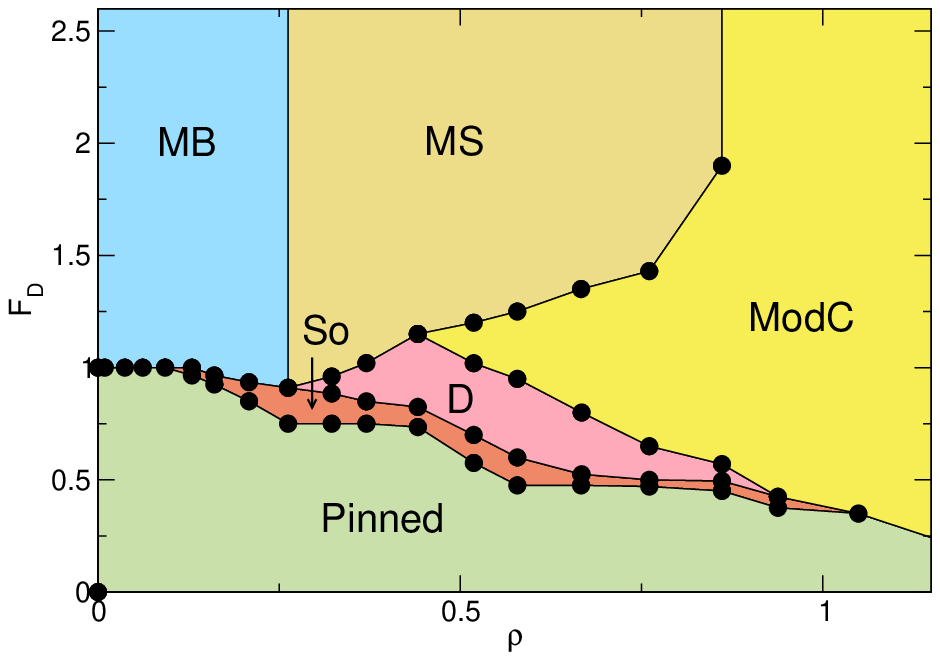}
\caption{Dynamic phase diagram as a function of $F_{D}$ vs $\rho$ for
a system with $B = 2.15$, $F_{p} = 1.0$, and $N_{p} = 8$.
The distinct pinned states are not resolved and are marked as
a Pinned region.
The moving states are soliton motion (So),
disordered (D) flow, moving bubble (MB),
moving stripe (MS), and moving modulated crystal (ModC).
}
\label{fig:23}
\end{figure}

From the features in the transport curves and
the particle configurations,
in Fig.~\ref{fig:23}
we construct a dynamic phase diagram
as a function of $F_{D}$ vs $\rho$ for the system
with $B = 2.15$, $F_{p} = 1.0$, and $N_{p} = 8$
where we highlight the pinned regime, moving soliton (So) state,
disordered motion (D) phase, moving
stripe (MS) state, moving bubble (MB)
phase, and moving modulated crystal (ModC).
The system cannot form a moving stripe when $\rho < 0.25$ and
instead enters the moving bubble phase.
The moving stripe phase occurs
in a window of density ranging from $0.25 \leq \rho < 0.8$.
In a portion of this density window, we find that the system first
passes through a moving modulated solid state before transitioning to
the moving stripe configuration.

\section{Reentrant Pinning Phases}

\begin{figure}
\includegraphics[width=\columnwidth]{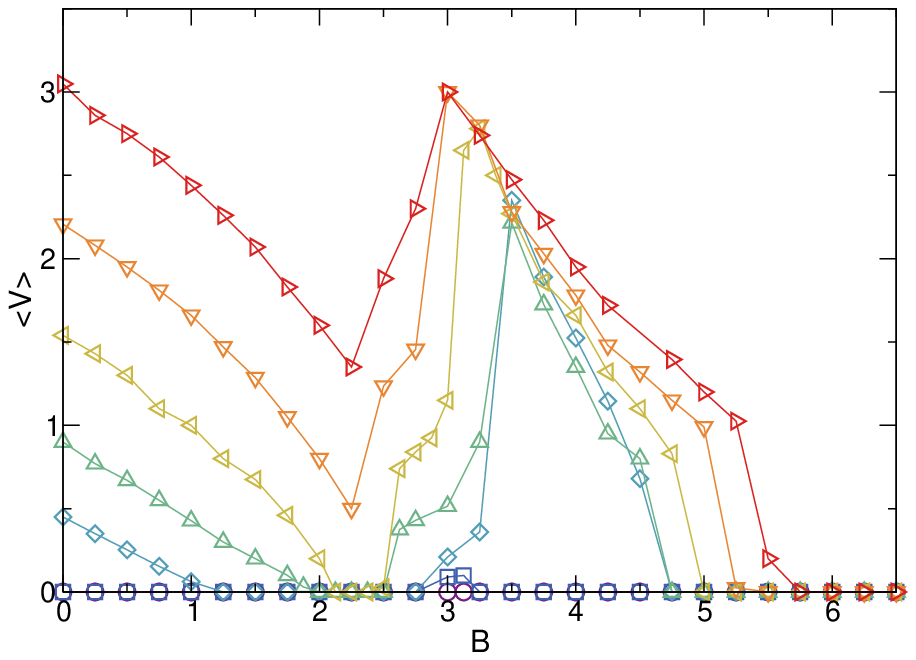}
\caption{$\langle V\rangle$ vs $B$ for a system with
$N_p = 8$, $F_{p} = 5.0$, and
$F_{D} = 4.0$ for
$\rho = 0.129$ (violet circles),
0.208 (dark blue squares),
0.262 (light blue diamonds),
0.322 (green up triangles),
0.44 (yellow left triangles),
0.67 (orange down triangles), and $1.16$ (red right triangles).
Two distinct pinned phases appear. Near $B=2.15$ is the pinned stripe
regime, and in the region above $B=4.75$ is the pinned small bubble
regime.
At very small $B$ is the weakly pinned crystalline state,
and there is a window of large bubble states in between the
stripe and small bubble states that is also weakly pinned.
As a result, there can be a doubly reentrant pinned state as a function
of increasing $B$.
}
\label{fig:24}
\end{figure}

We have demonstrated above that the stripe
state is the most strongly pinned phase;
however, for large $B$
the smaller bubbles are also very strongly pinned,
suggesting that it should be possible
to observe {\it multiple} reentrant pinning effects
under the right conditions.
We consider a sample with $N_{p} = 8$ and strong pinning
of $F_{p} = 5.0$ under a constant driving force of
$F_{D} = 4.0$.
In Fig.~\ref{fig:24}, we plot
$\langle V\rangle$ versus $B$ for this system
at $\rho = 0.129$, 0.208, 0.262, 0.322, 0.44, 0.67, and $1.16$.
For $\rho = 0.129$, the system is pinned for
all values of $B$, while for
$\rho = 0.262, 0.322$ and $0.44$,
the system is initially flowing at very small $B$, enters a
pinned stripe state
near $B = 2.15$,
and develops a finite velocity again
in the bubble phase for $B > 2.3$.
There is a local maximum in
the velocity 
near $B = 3.0$; however, as $B$ increases further,
the bubbles shrink in size
and become pinned again
for sufficiently large $B$,
reaching a second reentrant pinned state.
For $\rho > 0.44$, there is a dip in $\langle V\rangle$
near $B = 2.15$ in the stripe phase
but the system does not become reentrantly pinned;
this is followed by a local maximum in $\langle V\rangle$
in the large bubble phase
and a reentrant pinning of the small bubble state
for large $B$.
Going to large values of $B$ has an effect on the transport that is
similar to going to low particle density, since the increase of $B$ causes
the average spacing between adjacent bubbles to get larger as the
bubble radius becomes smaller. In particular, the depinning threshold
approaches $F_c/F_p=1.0$ for both large $B$ and small $\rho$. In
Fig.~\ref{fig:24}, we have fixed $F_D/F_p=0.8$, so once $F_c/F_p$ exceeds
this value either by decreasing $\rho$ or increasing $B$ in the small
bubble phase, the velocity drops to zero and a pinned state emerges.

\section{Discussion}
Pattern forming systems on a 1D substrate exhibit a variety of additional
effects that would be interesting to explore in future studies,
such as thermal or creep effects, where it would be possible
to compare stripe creep to bubble or anisotropic crystal creep, as well
as differences between creep in the elastic depinning regime and
creep in the plastic depinning regime.
Several previous studies of particles with purely
repulsive interactions coupled to a 1D periodic substrate demonstrated
reentrant melting or smectic phases as a function of
increasing substrate strength or filling
\cite{Chowdhury85,Chakrabarti95,Frey99,Radzihovsky01}.
In this work we considered purely dc driving,
but if ac driving were applied,
we would expect to observe Shapiro step phenomena
\cite{Reichhardt17}, and it would be possible to explore whether
the Shapiro steps are enhanced in the stripe phase compared to
the bubble phase.
Another direction would be to consider a 2D substrate
that could break apart the stripes or lock the flow of the 
stripe phase into particular
directions.
Other interesting effects to explore include
the effect of adding a small amount of random point disorder or
a random shift to the substrate.
Our results should be general to
the broader class of stripe or bubble-forming systems,
including those that
have different kinds of interactions, such as a purely repulsive
interaction potential with two length scales.

\section{Summary} 

We have numerically investigated the pinning and dynamics of
a two-dimensional pattern-forming system consisting
of particles with long-range repulsion and short-range
attraction interacting with a periodic one-dimensional 
substrate. In the absence of a substrate, this system forms a
crystal for very low attraction strength,
an anisotropic crystal at weak attraction strength,
a stripe lattice for intermediate attraction strength,
and a bubble lattice for strong attraction.
When a one-dimensional substrate is added to the sample,
we find that the stripe state is the most strongly
pinned overall, and is particularly strongly
pinned whenever the stripes are commensurate with
the substrate spacing and can align with the substrate minima.
In the bubble phase, when the bubbles are large they do not fit
into a single substrate minimum and are weakly pinned;
however, small bubbles that can fit inside the substrate minima
are strongly pinned.
The anisotropic crystal and stripe states
can depin plastically either via the motion
of solitons or directly into a disordered flow phase,
and at higher drives, the system can dynamically order into
a moving crystal state or
moving stripes that have rotated with respect to the pinned state
and are aligned with the driving direction.
We show that this system exhibits a wide variety of
dynamical phases, and that transitions
between the different phases are observable
as multiple steps or peaks in the velocity-force
curves and differential velocity
curves.
For small substrate lattice constants, the stripe and bubble phases are
replaced by
pinned modulated stripe phases with stripes
that are perpendicular to the substrate troughs, in contrast
to the case of large substrate spacing where the
stripes are aligned with the substrate troughs.
We map out the dynamic phases as a function of substrate strength,
attraction strength, density, and driving force.
At high densities, the stripes remain oriented with the substrate
trough direction and not with the driving direction even in the
moving state.
Our results are relevant for
a wide variety of similar pattern-forming systems in
both soft and hard matter systems
that are coupled to a periodic one-dimensional substrate.

\begin{acknowledgements}
We gratefully acknowledge the support of the U.S. Department of
Energy through the LANL/LDRD program for this work.
This work was supported by the US Department of Energy through
the Los Alamos National Laboratory.  Los Alamos National Laboratory is
operated by Triad National Security, LLC, for the National Nuclear Security
Administration of the U. S. Department of Energy (Contract No. 892333218NCA000001).
\end{acknowledgements}

\bibliography{mybib}

\end{document}